\documentclass[twocolumn]{article}
\usepackage[margin=0.6in,columnsep=0.15in]{geometry}
\usepackage[utf8]{inputenc}

\usepackage[usenames,dvipsnames]{xcolor}
\usepackage[colorlinks,citecolor=Gray,urlcolor=Gray,linkcolor=Blue]{hyperref}

\usepackage{graphicx}
\usepackage{mathtools}
\usepackage{booktabs}
\usepackage{tabularx}
\usepackage{tikz}
\usepackage{siunitx}
\usepackage{dcolumn}
\usepackage{comment}

\usepackage[lcgreekalpha,upint]{stix}

\usepackage[tiny]{titlesec}
\usepackage[inline]{enumitem}

\usepackage{algorithm} 
\usepackage{algpseudocode} 

\usepackage[normal]{caption}
\DeclareCaptionLabelSeparator{pipe}{ $|$ }
\usepackage{authblk}

\usepackage[square,sort&compress,comma,numbers]{natbib}

\usepackage{doi}

\usepackage{subfigure}

\usepackage[switch]{lineno}
\title{\Large Exploring Fractional Quantum Hall States with Deep Learning: From Abelian to Non-Abelian Phases}

\author[1,2$\ast,\dagger$]{Xiangyu Ma}
\author[1,3$\ast,\dagger$]{Kehuang Chen}
\author[1,4]{Yuntian Gu}
\author[1]{Weiluo Ren}
\author[5,6$\dagger$]{Ji Chen} 
\author[3,7$\dagger$]{Di Luo}
\author[2,8$\dagger$]{\\Fengcheng Wu}
\author[1$\dagger$]{Xiang Li}
\affil[1]{ByteDance Seed, Beijing 100098, China}
\affil[2]{School of Physics and Technology, Wuhan University, Wuhan 430072, China}
\affil[3]{Department of Physics, Tsinghua University, Beijing 100084, China}
\affil[4]{School of Intelligence Science and Technology, Peking University, Beijing 100871, China}
\affil[5]{School of Physics, Peking University, Beijing 100871, People's Republic of China}
\affil[6]{Interdisciplinary Institute of Light-Element Quantum Materials, Frontiers Science Center for Nano-Optoelectronics, Peking University, Beijing 100871, People's Republic of China}
\affil[7]{C.N. Yang Institute of Advanced Study, Tsinghua University, Beijing 100084, China}
\affil[8]{Wuhan Institute of Quantum Technology, Wuhan 430206, China}

\date{}

\makeatletter
\def\blfootnote{\xdef\@thefnmark{}\@footnotetext}
\makeatother

\begin{document}

\twocolumn[{%
  \maketitle
  \vspace{-3em}
  \begin{center}
  \begin{minipage}{0.85\linewidth}
    \small
    \paragraph{Abstract}

Fractional quantum Hall systems host strongly correlated topological phases with potential applications in topological quantum computation, yet identifying their microscopic ground states remains challenging because competing phases are often nearly degenerate and Landau-level mixing is difficult to treat. Here, we introduce a symmetry-adapted neural-network wavefunction that describes Abelian and non-Abelian states while incorporating all Landau levels. The method accurately captures composite-fermion liquids and the emergence of Wigner crystallization at low filling. When the first Landau level is partially filled, finite-size overlaps and entanglement spectra provide evidence consistent with $\bar{2}^{k}1^{k+1}$-type parton phases at $\nu=4/3$ and $7/5$, while the $\nu=3/2$ state exhibits anti-Pfaffian-like entanglement structure. These results demonstrate that neural-network wavefunctions offer a flexible framework for studying competing topological orders in fractional quantum Hall systems.

  \end{minipage}
  \end{center}
  \vspace{1em}
}]
\blfootnote{$^{*}$These two authors contributed equally.}
\blfootnote{{\raggedright\hangindent=2.2em\hangafter=1\textsuperscript{$\dagger$} Email: {\urlstyle{same}
\nolinkurl{mxy12138@whu.edu.cn},
\nolinkurl{ckh24@mails.tsinghua.edu.cn},
\nolinkurl{ji.chen@pku.edu.cn}, 
\nolinkurl{diluo@tsinghua.edu.cn}, 
\nolinkurl{wufcheng@whu.edu.cn}, \nolinkurl{lixiang.62770689@bytedance.com}\par}}}


\let\oldAA\AA
\renewcommand{\AA}{\text{\normalfont\oldAA}}
\newcommand{\red}[1]{\textcolor{red}{#1}}
\newcommand{\bx}{\mathbf{x}}
\newcommand{\br}{\mathbf{r}}
\newcommand{\bk}{\mathbf{k}}
\newcommand{\bR}{\mathbf{R}}
\newcommand{\bL}{\mathbf{L}}
\newcommand{\bg}{\mathbf{g}}
\newcommand{\bG}{\mathbf{G}}
\newcommand{\bom}{\mathbf{\Omega}}

\hyphenation{Schrö-din-ger}

\section{Introduction}
Two-dimensional electron systems subjected to strong perpendicular magnetic fields exhibit the fractional quantum Hall effect (FQHE) at partial Landau-level fillings~\cite{Tsui1982,Laughlin1983}. These strongly correlated phases are characterized by precisely quantized Hall conductivity and dissipationless electric transport. The recent observation of analogous states in moir\'e materials has further revitalized the field by extending fractional quantum Hall physics beyond conventional Landau levels to highly tunable quantum platforms~\cite{Cai2023,Zeng2023,Park2023,Xu2023,Lu2024}.

The FQHE encompasses a hierarchy of phases whose character depends sensitively on Landau-level filling, electron-electron interactions, and Landau-level mixing. In the lowest Landau level, many experimentally observed fractions are successfully described by Laughlin \cite{Laughlin1983} and composite-fermion (CF) theory~\cite{Jain1989,Jain2007}, in which quasiparticles in these states can carry fractional electric charge and obey Abelian exchange statistics~\cite{Laughlin1983,Halperin1984}. At higher Landau-level fillings, however, interactions can stabilize more intricate paired or clustered phases whose quasiparticles may exhibit non-Abelian statistics~\cite{Willett1987,Moore1991,ReadRezayi1999}. Braiding these quasiparticles performs transformations within a degenerate quantum-state manifold, making such phases attractive platforms for topological quantum computation~\cite{Nayak2008}.

Despite substantial theoretical and experimental progress, a systematic understanding of FQHE phases across different filling factors remains elusive. Over the decades, a variety of ingenious trial wave functions have been developed, providing profound insights into the nature of these strongly correlated states. However, when Landau-level mixing becomes significant, an accurate description of the many-body physics becomes considerably more challenging. This challenge is especially acute in higher Landau levels, where several competing candidate states can have very close energies, leaving the identity of the ground state an open question. Numerical methods such as exact diagonalization (ED)~\cite{Haldane1983,Morf1998,RezayiHaldane2000} and the density matrix renormalization group (DMRG)~\cite{White1992,Feiguin2008,Zaletel2015} have been instrumental in advancing our understanding, while their applicability remains limited by accessible system sizes, geometries, and the number of Landau levels that can be retained.

Neural-network quantum states have recently emerged as a powerful framework for studying strongly correlated quantum matter. Owing to their expressive representations of many-body correlations, they have achieved competitive accuracy in problems involving models, molecules, solids and topological phases~\cite{Carleo2017,luo2019,Pfau2020,Hermann2020,Li2022,hegnet,ferminet_gas,valenti2025,ueg2024,ti2025}. Existing neural-network studies of the FQHE~\cite{Teng2025,Qian2025,Gattu2025,zhu2026,pf2026}, however, have largely focused on a limited set of filling factors and predominantly Abelian states. Whether these methods can provide a unified and quantitatively reliable description across a broader range of fillings, especially for non-Abelian phases and substantial Landau-level mixing, remains an important open question.

\begin{figure*}[t]
    \centering
    \includegraphics[width=1.8\columnwidth]{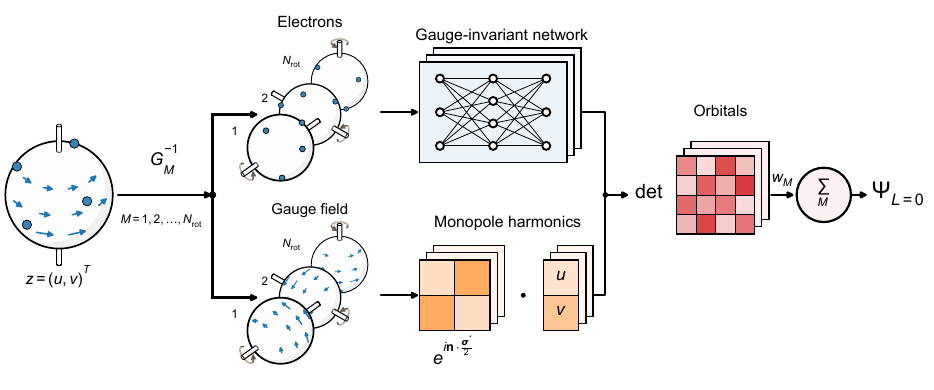}
	\caption{\textbf{Neural-network architecture.} Haldane spinors are used as inputs to the neural network. The upper channel processes the electron coordinates through the neural network, while the lower channel generates the gauge phase associated with the transformation of the vector potential. To achieve rotational invariance, the two contributions are combined to obtain the wave-function amplitude under each $SO(3)$ rotation, and the group integral is approximated by numerical quadrature summation.
}
    \label{fig:1}
\end{figure*}

In this work, we develop a symmetry-adapted neural-network wave function for fractional quantum Hall systems on the Haldane sphere. The ansatz combines gauge-invariant many-body features with monopole harmonics and restores rotational symmetry through numerical $SO(3)$ integral, while its non-holomorphic structure naturally incorporates Landau-level mixing. In the lowest Landau level, the optimized network largely retains the familiar Laughlin and composite-fermion states, before evolving toward crystal at very low fillings. A qualitatively different picture emerges when the first Landau level (1LL) is partially occupied: the modified interactions drive the system away from conventional Abelian states and allow exotic non-Abelian states to develop. This contrast demonstrates that the neural network can capture changes in topological order across Landau levels without prescribing the underlying phase in advance.
\section{Results}  

\subsection{Simulation workflow}

We consider quantum Hall systems in spherical geometry, with
electrons confined to the sphere surface \cite{sphere}. 
A magnetic monopole of strength $Q$, placed at the center,
generates a radial magnetic field through the sphere.
The sphere radius is
$R=\sqrt{Q}\ell_B$, where $\ell_B=\sqrt{\hbar c/eB}$ is the magnetic length and $B$ is the magnetic field.
The many-body Hamiltonian reads
\begin{equation}
\hat{H}
=
\frac{\ell_B^{2}\hbar\omega_c}{2R^{2}}
\sum_i
\left|
\hat{\boldsymbol{\Lambda}}_i
\right|^{2}
+
\frac{e^{2}}{\epsilon}
\sum_{i<j}
\frac{1}{
|\mathbf{r}_i-\mathbf{r}_j|
}\ ,
\label{eq:spherical_hamiltonian}
\end{equation}
where $\omega_c=eB/(mc)$ is the cyclotron frequency. For electrons of charge $-e$,
$\hat{\boldsymbol\pi}_i=-\mathrm{i}\hbar\nabla_{\mathbf r_i}+(e/c)\mathbf A(\mathbf r_i)$
is the gauge-covariant momentum, and
$\hat{\boldsymbol\Lambda}_i=(\mathbf r_i\times\hat{\boldsymbol\pi}_i)/\hbar$
is the dimensionless tangential angular momentum.
The Landau-level mixing parameter is
$\kappa=(e^2/\epsilon\ell_B)/(\hbar\omega_c)$.


The network architecture is illustrated in Fig.~\ref{fig:1}.  Each electron is represented by the Haldane spinor
$
z=(u,v)^{\mathsf T}
=
\left(
\cos\frac{\theta}{2}e^{\mathrm{i}\phi/2},
\sin\frac{\theta}{2}e^{-\mathrm{i}\phi/2}
\right)^{\mathsf T}
$
which is processed through two complementary channels.
The upper channel maps the spinor to the unit-sphere direction
$\mathbf n=z^\dagger\boldsymbol{\sigma}^*z$, with physical position
$\mathbf r=R\mathbf n$, and feeds these coordinates into the PsiFormer
network \cite{Psiformer,Qian2025}. This channel generates non-holomorphic correlations and therefore allows the variational state to incorporate Landau-level mixing. The lower one constructs magnetic spherical harmonics
$u^{Q+m}v^{Q-m}$ from the spinors, compensating the gauge singularity induced by the vector potential $\mathbf{A}$.
The outputs of the two channels are combined to form collective orbitals, whose Slater determinant gives the final wavefunction.

Moreover, the incompressible FQH liquids are angular momentum
singlets,
whereas a generic neural wavefunction may contain
components with $L>0$.  
To impose this symmetry, we approximate the singlet state by numerical quadrature summation over the $SO(3)$ group:
\begin{equation}
\begin{aligned}
\langle X|\psi_{L=0}^{\rm net}\rangle
&=
\int_{SO(3)} dM\ 
\langle X|\hat{U}_Q(M)|\psi^{\rm net}\rangle\\
&\approx\sum_M w_M\ \langle X|\hat{U}_Q(M)|\psi^{\rm net}\rangle\ ,
\label{eq:L0_projection}
\end{aligned}
\end{equation}
where $dM$ is the normalized Haar measure, $\hat{U}_Q(M)$ is the physical rotation, $X=(\mathbf r_1,\ldots,\mathbf r_N)$, and $N$ is the electron number.

In a fixed gauge, a spatial rotation generally changes the vector potential by a gauge transformation, giving rise to a configuration-dependent phase $\chi_M^{\mathrm{tot}}(X)$. Since the magnetic field itself is rotationally invariant, this phase can be incorporated naturally through the $SU(2)$ action on the Haldane spinors. For each rotation $M$, the rotated wave function can be written as
\begin{equation}
\begin{aligned}
\langle X|\hat U_Q(M)|\psi^{\mathrm{net}}\rangle
&=e^{i\chi_M^{\mathrm{tot}}(X)}\psi^{\rm net}(M^{-1}X)\\
&= \psi^{\mathrm{net}}
\left(
\{G_M^{-1}z_i\}_{i=1}^{N}
\right).
\end{aligned}
\label{eq:covariant_spinor_rotation}
\end{equation}
where $G_M$ is the $SU(2)$ representation of $M$. Finally, the network parameters are optimized using variational Monte Carlo by minimizing the energy expectation value, yielding a variational approximation to the many-body ground state.

\subsection{Abelian states and Wigner crystals} 
\begin{figure*}[t]
    \centering
    \includegraphics[width=\textwidth]{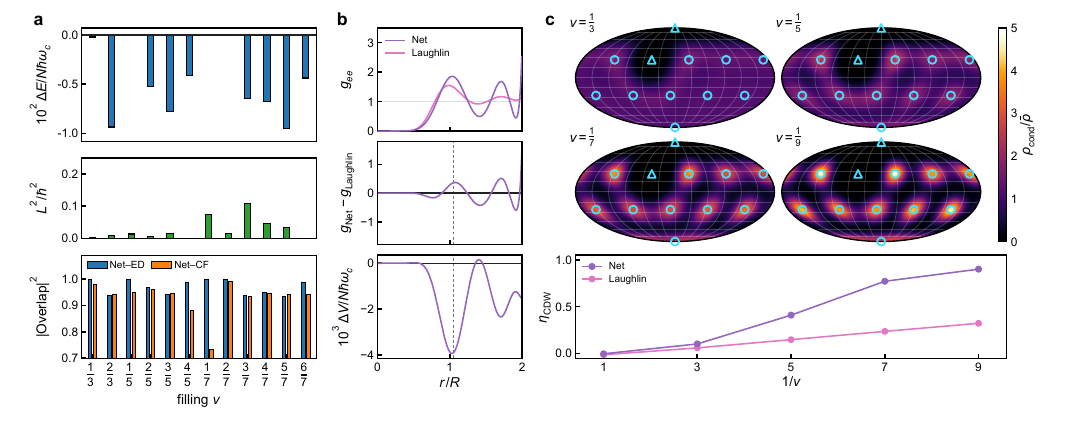}
    \caption{\textbf{Abelian states.} Our results are labeled as Net. All systems are at $\kappa=1$.
    ${\rm \mathbf{a}}$, energy difference $\Delta E=E_{\rm Net}-E_{\rm ED}$ per particle relative to truncated ED, total angular momentum of Net, and squared overlaps of Net with ED and with the
    CF trial state; the
    $\nu=1/7$ system here is $N=6$.
    ${\rm \mathbf{b}}$, pair correlation of Net and the Laughlin state at $\nu=1/7$ for $N=12$ (top),
    their difference $g_{\rm Net}-g_{\rm Laughlin}$ (middle), and the interaction-energy difference
    $\Delta V=V_{\rm Net}-V_{\rm Laughlin}$ accumulated up to separation $r$ (bottom). The dashed line marks the
    nearest-neighbor chord distance of the $N=12$ Thomson icosahedron.
    ${\rm \mathbf{c}}$, conditional density for $N=12$ at different fillings $\nu=1/m$.
    One electron is pinned at the north pole and the second at the first-peak separation of the pair correlation function (pinned positions marked by triangles), in units of its mean $\bar\rho=(N-2)/4\pi R^2$. The remaining ten vertices of the reference Thomson icosahedron are marked by circles (top), and the corresponding CDW order parameter $\eta_{\rm CDW}$ is compared with the Laughlin state (bottom).
    }
    \label{fig:2}
\end{figure*}

Abelian fractional quantum Hall phases are characterized by quasiparticles that change the many-body wave
function only by an overall phase when braided. The best-known examples are the Laughlin states at filling factor
$\nu=1/(2n+1)$~\cite{Laughlin1983}, while composite-fermion theory extends this framework to the broader Jain
sequence and its particle--hole conjugates~\cite{Jain1989,Jain2007}. Their well-established microscopic
descriptions make it possible to assess the neural-network ansatz systematically as the filling factor varies.

We first benchmark the ansatz across fluxes spanning the Jain sequence within $\nu<1$.
As shown in Fig.~\ref{fig:2}a, our calculated energy is lower than the largest ED calculation feasible for each studied $\nu$. The optimized states remain close to rotational singlets and retain strong fidelity with
the corresponding trial states, except at $\nu=1/7$, where the ground state departs from the
Laughlin liquid. Thus, over most of the Jain sequence, Landau-level mixing dresses an otherwise
intact CF states rather than destroying its topological order.

To elucidate the microscopic origin of the energy improvement, we calculate the electron-electron pair correlation function $g_{ee}(r)$, where $r=2R\sin(\theta/2)$ is the chord distance between two electrons on the sphere. The Coulomb interaction energy is an exact linear functional of $g_{ee}(r)$,
\begin{equation}
  \frac{V(r)}{N}
  =
  \frac{N e^2}{4\epsilon R^2}
  \int_0^{r}g_{ee}(r')\,dr'.
  \label{eq:Eg}
\end{equation}
As shown in Fig.~\ref{fig:2}b for $N=12$ particles at $\nu=1/7$, the system under Landau-level mixing exhibits a broader Pauli hole and more pronounced correlation shells, indicating enhanced spatial organization that lowers the Coulomb interaction energy.

This emerging shell structure suggests the transition from an FQH liquid to a Wigner crystal---or charge-density wave (CDW), in which electrons localize into an ordered spatial pattern at low filling factors~\cite{Lam1984,Yang2001,Zuo2020,Zhao2018}. Locating this transition is usually challenging because conventional variational studies compare separately constructed liquid and crystal trial states, potentially biasing the result toward the assumed phases. Our neural-network ansatz instead allows liquid and crystalline correlations to compete within a single variational framework, without presupposing the nature of the optimized state.

\begin{figure*}[t]
    \centering
    \includegraphics[width=\textwidth]{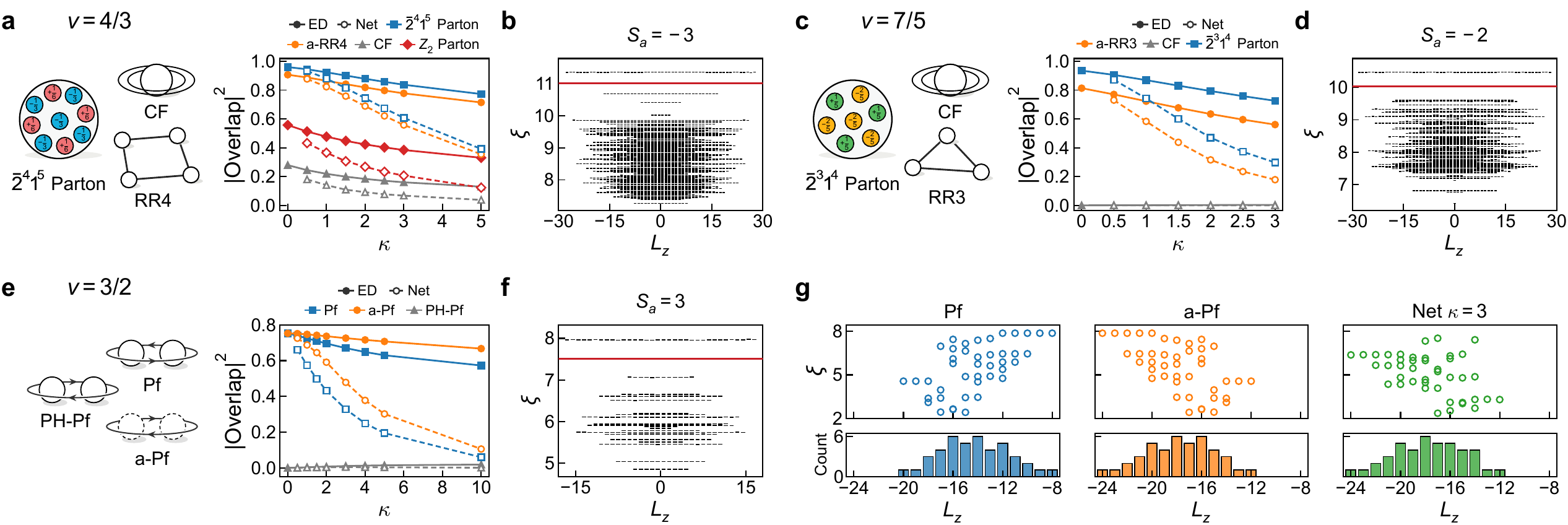}
    \caption{
    \textbf{Non-Abelian states.} Neural-network results compared with two-Landau-level ED. $S_a=N_a/\nu_a-2(Q+1)$ denotes the active-shell shift in the 1LL.
    \textbf{a}, candidate parton, CF, and RR states at $\nu=4/3$, with squared overlaps shown in the right panel. Results use $(N,2Q)=(20,13)$  for the CF and $Z_2$ parton states and $(N,2Q)=(18,13)$ for the a-RR4 and $\bar{2}^{4}1^{5}$ states.
    \textbf{b}, PES of the leading network state at $\nu=4/3$ ($\kappa=1$) at $(N,2Q)=(18,13)$ , obtained by tracing out $N-N_A$ particles with $N_A=5$. The red line marks the entanglement gap.
    \textbf{c}, candidate states and squared overlaps at $\nu=7/5$, using $(N,2Q)=(16,9)$  for the CF state and $(N,2Q)=(22,15)$  for the a-RR3 and $\bar{2}^{3}1^{4}$ states.
    \textbf{d}, PES of the leading network state at $\nu=7/5$ ($\kappa=1$) at $(N,2Q)=(22,15)$  with $N_A=4$.
    \textbf{e}, Pf, a-Pf, and PH-Pf candidates at $\nu=3/2$, using $(N,2Q)=(20,11)$, $(18,11)$ , and $(22,13)$, respectively.
    \textbf{f}, PES of the Pf network state at $\nu=3/2$ ($\kappa=1$) at $(N,2Q)=(20,11)$  with $N_A=3$.
    \textbf{g}, upper panel plots OES of the four-quasihole Pf, its particle--hole conjugate, and the network state at the particle-hole-symmetric shift $(N,2Q)=(22,13)$ and $\kappa=3$, with $N_A=4$. Four-quasihole Pf reference is the $L=0$ zero mode of the three-body parent Hamiltonian with the lowest Coulomb energy; the network state is projected onto the sector with a filled lowest Landau level. Lower panel counts the low-lying states in each $L_z$ sector.
    }
    \label{fig:3}
\end{figure*}

To visualize the crystal order in real space, we fix the global $SO(3)$ orientation of the wavefunction by pinning one electron at the north pole and the second at the first peak of the pair correlation function. Fig.~\ref{fig:2}c shows the calculated conditional electron density. For the specific case $N=12$, its density maximum closely match the corresponding Thomson configuration which forms a icosahedral geometry and minimizes the interaction energy~\cite{Thomson1904}.  We define the CDW order parameter following Refs.~\cite{Blakie2020,MendozaCoto2022},
\begin{equation}
  \eta_{\rm CDW}=\frac{\rho_{\rm v}-\rho_{\rm f}}{\rho_{\rm v}+\rho_{\rm f}},
  \label{eq:eta_cdw}
\end{equation}
where $\rho_{\rm v}$ and $\rho_{\rm f}$ denote the mean conditional densities at the remaining icosahedral vertices and interstitial face centers, respectively.  By construction, $\eta_{\rm CDW}\in[0,1]$ vanishes for a uniform liquid and approaches unity for a localized crystal. At $\nu=1/3$, the conditional density away from the pinned electrons remains nearly uniform with a small $\eta_{\rm CDW}$ close to the Laughlin baseline, characteristic of an incompressible quantum liquid. As $\nu$ decreases to $1/7$ and $1/9$, $\eta_{\rm CDW}$ increases sharply above the Laughlin reference as pronounced peaks develop on the icosahedral vertices, providing a direct real-space signature of Wigner crystallization.

\subsection{Non-Abelian states}

FQHE phenomenon has been observed over a broad range of filling factors, including in higher Landau levels.
Although one might expect the underlying physics to remain broadly similar across Landau levels, distinct Landau level wavefunctions modify the effective interactions between electrons and can stabilize qualitatively different many-body phases. In the first Landau level, the competition among candidate phases makes identifying the ground state a nontrivial challenge. Addressing this challenge requires accurate and reliable numerical methods capable of distinguishing between competing possibilities.

We begin with the filling factor $\nu=4/3$ (here we consider a fully spin-polarized system), which might be expected to exhibit a Laughlin-like state. However, the overlaps with our neural-network ground state suggest a different picture (see Fig.~\ref{fig:3}a,b). The optimized network has relatively small overlaps with the Laughlin state and the $Z_n$ ($n\bar{n}111$) parton states for $n=1,2$, all of which host Abelian anyons~\cite{Balram2020,Faugno2021}. Note that the $Z_3$ and $Z_4$ anyon states require system sizes beyond the reach of our calculations and are therefore not considered here. In contrast, substantially larger overlaps are found with the $\bar{2}^{4}1^{5}$ parton state and the anti-Read--Rezayi $k=4$ (a-RR4) state, which have similar non-Abelian topological order~\cite{Balram2019}. This result is further supported by the particle entanglement spectrum (PES) in Fig.~\ref{fig:3}b, whose low-lying levels reproduce the characteristic a-RR4 quasihole counting, whereas the spectrum at the Laughlin shift does not exhibit the expected Laughlin counting (Supplementary Fig.~3). Moreover, the finite-size analysis shows that these overlap orders persist up to $N=34$ system size (Supplementary Fig.~2), supporting a non-Abelian ground state at $\nu = 4/3$.

At $\nu=7/5$, a clearer picture emerges (see Fig.~\ref{fig:3}c,d). The $\bar{2}^{3}1^{4}$ parton state provides the closest match to our neural-network ground state, followed by a-RR3, with the CF state showing nearly vanishing overlap. This hierarchy persists as Landau-level mixing grows. Despite their distinct constructions, the $\bar{2}^{3}1^{4}$ parton and a-RR3 states host the same Fibonacci anyons~\cite{Balram2019}. Building on previous numerical evidence for the a-RR3 phase at the corresponding $\nu=12/5$ filling~\cite{Zhu2015,Mong2017}, our results point more specifically to the $\bar{2}^{3}1^{4}$ parton wave function as the leading ground-state candidate within this phase. 

At $\nu=3/2$, the contrast to $\nu=1/2$ is again sharp. The half-filled lowest Landau level  is a gapless composite-fermion
liquid~\cite{HalperinLeeRead1993}, but in the first Landau level, the weaker short-range repulsion can instead drive composite-fermion pairing at $\nu=3/2$, which is relevant to the experimentally observed FQHE at  
$\nu=5/2$~\cite{Willett1987}. The leading
candidates are the Moore--Read Pfaffian (Pf)~\cite{Moore1991,Nayak2008},
its particle--hole conjugate [i.e., the anti-Pfaffian (a-Pf)]~\cite{Lee2007,Levin2007}, and the particle--hole-symmetric Pfaffian (PH-Pf)~\cite{Son2015}. When the Hamiltonian is projected to the 1LL,
ED~\cite{Morf1998,RezayiHaldane2000} and DMRG~\cite{Feiguin2008} studies support
this pairing picture. However, the exact particle--hole symmetry of a single Landau level leaves Pf and a-Pf degenerate.
Lifting this degeneracy requires Landau-level mixing: multi-Landau-level infinite DMRG~\cite{Zaletel2015} and
perturbative ED including higher-order pseudopotentials~\cite{Rezayi2017} favor the anti-Pf,
whereas studies incorporating finite layer width favor the Pf~\cite{Pakrouski2015}. By contrast, the
PH-Pf is energetically disfavored for the pure Coulomb interaction~\cite{RezayiPakrouskiHaldane2021}.

The calculated overlaps between our network and trial states are plotted in Fig.~\ref{fig:3}e. At their respective shifts and system sizes, the optimized states show clear agreement with the Pf (a-Pf) states, which is further confirmed by the calculated PES in Fig.~\ref{fig:3}f. Our results support the Pf and a-Pf states, while the PH-Pf state is essentially absent. To examine the competition between the Pf and a-Pf states more directly, we consider the particle--hole-symmetric shift, where both candidates compete within the same Hilbert space.
For reference, the orbital entanglement spectra (OES)~\cite{LiHaldane2008} of Pf and a-Pf at this shift are
mirror images, reflecting opposite edge chirality. By comparison, the OES of the optimized state follows that of the
a-Pf as shown in Fig.~\ref{fig:3}g. Landau-level mixing thus favors the
a-Pf~\cite{Zaletel2015,Rezayi2017}.

Our results also highlight the sensitivity of non-Abelian states to Landau-level mixing. Restricting calculations to a few Landau levels omits virtual excitations into higher levels, making this approximation less reliable at experimentally relevant mixing strengths, typically $\kappa\sim1-3$. As shown in Fig.~\ref{fig:3}, the overlaps between our optimized states and the corresponding topological trial states fall more rapidly with increasing $\kappa$ than projected ED predicts, especially at $\nu = 3/2$. Notably, Although the full overlap including all Landau levels approaches zero at large $\kappa$ , the Pf PES, evaluated after projecting the state onto the active 1LL, retains a clear Moore--Read entanglement gap up to $\kappa=10$ (Supplementary Note~5). This contrast between the rapid collapse of the full-space overlap and the persistence of the projected entanglement gap highlights a key limitation of Landau-level projection: omitting higher-Landau-level leakage can substantially overestimate the stability of the topological phase. Consequently, projected approaches that neglect full Landau-level mixing must be interpreted with caution, and non-Abelian phases under realistic experimental conditions may be far more fragile against $\kappa$ than projected calculations suggest. Stronger magnetic fields, which lower $\kappa$ in conventional semiconductor systems, may therefore be essential to stabilize them.

\section{Discussion}
Over the past several decades, physically motivated trial wavefunctions have shaped our understanding of fractional quantum Hall systems, where strong correlations make exact solutions inaccessible in most cases. As attention turns to more complex phases and substantial Landau-level mixing, identifying suitable trial states becomes increasingly difficult. Deep learning offers a promising way to explore this challenging landscape. Our results show that a single neural-network wavefunction ansatz can describe both Abelian and non-Abelian states while incorporating contributions from all Landau levels without an explicit cutoff or perturbative treatment of mixing. This allows electrons to reduce their Coulomb repulsion by admixing higher Landau levels at a finite kinetic energy cost, a balance that cannot be captured within a strictly projected Hamiltonian.

Based on this method, several important open questions may become accessible. One particularly intriguing problem concerns the graviton mode recently observed in optical experiments \cite{graviton}, whose microscopic nature and relation to the geometric response of FQH states are not yet fully understood. Another is the characterization of other even-denominator phases \cite{nsr_exp}, where competing topological orders remain difficult to distinguish. Addressing these problems will require wavefunction ansatz that is both more expressive and computationally efficient in order to capture ground as well as excited states. Incorporating rotational symmetry directly into the network architecture may therefore be an important step toward resolving these open questions. Ultimately, continued advances in neural-network wavefunctions may open a practical route toward resolving the most challenging open questions in fractional quantum Hall physics.

\section{Methods}

\paragraph{Neural-network architecture.}
We use a PsiFormer backbone~\cite{Psiformer,Qian2025} to construct complex-valued functions
$f_{im}(\mathbf r_j;\mathbf r_{\neq j})$,
which are combined with lowest-Landau-level monopole
harmonics to form the orbital matrix.
The resulting neural-network wave function takes the form
\begin{equation}
\begin{aligned}
\psi^{\rm net}(X)
&=
e^{J(X)}
\det\!\left[\phi_i\!\left(z_j;z_{\neq j}\right)\right],\\
\phi_i\!\left(z_j;z_{\neq j}\right)
&=
\sum_{m=-Q}^{Q}
f_{im}\!\left(\mathbf r_j;\mathbf r_{\neq j}\right)
u_j^{Q+m}v_j^{Q-m}.
\end{aligned}
\label{eq:net_ansatz}
\end{equation}
Here $J(X)$ is the Jastrow factor, $Q$ is the monopole
strength, and $u_j^{Q+m}v_j^{Q-m}$ are the unnormalized
lowest Landau-level monopole harmonics.

\paragraph{Gauge-covariant rotations.}
Rotations on the Haldane sphere require additional care due to
the magnetic monopole. Consider an ordinary spatial rotation
$\hat U_S(M)$ acting as
$\hat{U}_S(M)^\dagger \hat{H}(\hat{\mathbf{r}})\hat{U}_S(M)=\hat{H}(M\mathbf{\hat{r}})$.
Under this rotation, the Hamiltonian is physically invariant,
but the vector potential transforms as
\begin{equation}
\mathbf A(\mathbf r)\longrightarrow
\mathbf A_M(\mathbf r)
=
M^{-1}\mathbf A(M\mathbf r).
\end{equation}
Since the magnetic field is rotation-invariant,
these two vector potentials differ only by a pure gauge
transformation, locally satisfying
\begin{equation}
\frac{e}{c}(\mathbf A_{M^{-1}}-\mathbf A)
=
2\hbar Q\nabla\chi_M.
\end{equation}

Hamiltonian is therefore symmetric for physical rotation
$\hat U_Q(M)=e^{i2Q\chi_M(\mathbf{\hat{r}})}
\hat U_S(M)$, which acts on the wave function as
\begin{equation}
\langle\mathbf{r}|\hat{U}_Q(M)|\psi\rangle
=
e^{i 2Q\chi_M({\mathbf r})}
\psi(M^{-1}{\mathbf r}).
\label{eq:gauge_rotation}
\end{equation}

A spatial rotation \(M\in SO(3)\) can be represented by a corresponding \(SU(2)\) matrix \(G_M\), which acts on the Haldane spinor according to the spin-\(1/2\) representation as 
\begin{equation}
\langle \mathbf{r}|\hat{U}_{1/2}(M)|z\rangle
=
G_M^{-1}z({\mathbf r})=e^{i\chi_M(\mathbf r)}
z(M^{-1}\mathbf r)\ ,
\end{equation}
where $G_M=e^{i\mathbf{n}_M\cdot\frac{\mathbf{\sigma^*}}{2}}$. Based on this spinor, we can construct both the gauge-invariant position $\mathbf{n}=z^\dagger\mathbf{\sigma}^*z$ and lowest Landau-level functions. Since the constructed wavefunction is a product of gauge-invariant functions $f(\mathbf{r}_i;\mathbf{r}_{\neq j})$ and lowest Landau-level functions, its transformation under a rotation can be written as
\begin{align}
&\langle X|\hat U_Q(M)|\psi^{\mathrm{net}}\rangle
\nonumber\\
&\quad=
\left[
\prod_{i=1}^{N}
e^{i2Q\chi_M({\mathbf r}_i)}
\right]
\psi^{\rm net}
\left(
\{z(M^{-1}{\mathbf r}_i)\}_{i=1}^{N}
\right)
\nonumber\\
&\quad=
\left[
\prod_{i=1}^{N}
e^{i2Q\chi_M({\mathbf r}_i)}
\right]
\psi^{\rm net}
\left(
\{e^{-i\chi_M(\mathbf{r}_i)}G_M^{-1}z_i\}_{i=1}^{N}
\right)
\nonumber\\
&\quad=
\psi^{\rm net}
\left(
\{G_M^{-1}z_i\}_{i=1}^{N}
\right).
\label{eq:covariant_spinor_rotation_2}
\end{align}
Thus, the gauge phase required by the physical rotation
is automatically incorporated through the spinor
transformation, allowing gauge-covariant rotations to
be implemented simply by applying $G_M^{-1}$ to every spinor.

\paragraph{Numerical rotational projection.}
The continuous $L=0$ projection is evaluated numerically by replacing the
normalized Haar integral over $SO(3)$ with a finite weighted sum,
\begin{equation}
\hat P_{L=0}
=
\int_{SO(3)}dM\,\hat U_Q(M)
\simeq
\sum_M w_M\hat U_Q(M).
\end{equation}
For a quadrature with degree of exactness $d$, the
Wigner-$D$ matrices satisfy
\begin{equation}
\sum_M w_M D^L_{mn}(M)=0,
\qquad 1\le L\le d,
\end{equation}
so that the corresponding nonzero-angular-momentum components cancel in the
discrete rotational average.
Common choices include product Euler-angle quadratures, Lebedev-based
quadratures \cite{ShimizuTsunoda2023,GrafPotts2009}, and optimized cubature rules directly on $SO(3)$
\cite{Graf2013,ShimizuTsunoda2023}.

\paragraph{Workflow and computational details.}
This work was developed on open-sourced \texttt{jaqmc} package on Github
(\url{https://github.com/bytedance/jaqmc}).
The parameters of the network wavefunction were optimized
using MARCH optimizer \cite{gu_solving_2025}. Energy and topological observables were obtained from Monte Carlo evaluation. 
ED benchmarks were computed independently with \texttt{diagHam}.
Network and optimization hyperparameters are listed in Supplementary Table~1.

\section*{Data availability}
All data generated or analyzed during this study are included in this published article (and its supplementary information files).

\paragraph{Note added} A recent preprint has reported related results~\cite{magnet}.

\begingroup
\setlength\bibsep{0pt}
\newcommand{\mathsl}{\mathit}
\footnotesize
\bibliography{ref}
\endgroup

\subsection*{Acknowledgments}
\begingroup
\footnotesize
The authors thank Tongzhou Zhao for helpful discussions. We thank ByteDance Seed group for inspiration and encouragement. 
X. M. was supported by the Basic Research Program for Undergraduate Students of the National Natural Science Foundation of China under Grant No. 126B1010.
F. W. was supported by the National Key Research and Development Program of China (Grants No. 2021YFA1401300 and No. 2022YFA1402400), and the National Natural Science Foundation of China (Grants No. 12274333 and No. 12550404).
J. C. was supported by the National Natural Science Foundation of China under Grant No. 12334003. D. L. was supported by the Beijing Major Science and Technology Project under Contract No. Z251100008125036.

\endgroup

\subsection*{Author contributions}

\begingroup
\footnotesize
X. M., K. C., F. W. and X. L. conceived the study; X. M. and K. C. developed the method, performed implementations, simulations, and data analyses; J. C., D. L., F. W. and X. L. supervised the project. All authors wrote the paper.
\endgroup

\subsection*{Competing interests}
\begingroup
\footnotesize
The authors declare no competing interests.
\endgroup

%

\end{document}


\title{Supplementary information: Exploring Fractional Quantum Hall States with Deep Learning: From Abelian to
Non-Abelian Phases}

\maketitle
\section{Hyperparameters for simulations}

The recommended hyperparameters are listed in Supplementary Table~\ref{tab:ge_hyper}. 
\begin{table}[htb]
\caption{\textbf{General architecture and optimization hyperparameters.}
}
\centering
\small
\begin{tabular}{lclc}
\toprule
Hyperparameter & Value & Hyperparameter & Value \\
\midrule
One-electron feature dimension & 256
    & Number of determinants & 1 \\
Number of PsiFormer layers & 4
    & Attention heads / dimension per head & $4/64$ \\
Optimizer & MARCH
    & Initial learning rate & $0.05$ \\
Maximum SR update norm & $0.1$
    & SR damping & 1e-3 \\
MARCH momentum coefficient $\mu$ & 0.9
    & MARCH decay coefficient $\beta$ & 0.5 \\
MARCH mode & Variance
    & Maximum SR condition number & 1e7 \\
Batch size & 1024
    & Number of optimization steps & 2e4 \\
Gradient-estimator clipping & IQR (scale 100)
    & MCMC steps per optimization step & 10 \\
MCMC burn-in steps & $100$
    & Target MCMC acceptance & 50--55\% \\
Initial MCMC move width & $0.1$
    & Learning-rate delay / decay & $2000/1.0$ \\
Numerical precision & Float32
    & SR linear-algebra precision & Float64 \\
\bottomrule
\end{tabular}
\label{tab:ge_hyper}
\end{table}
%


\section{Calculated results}

Supplementary Tables~\ref{tab:jain}--\ref{tab:nu75} list the energies,
angular momenta and overlaps of all systems shown in the main text. The overlap between the neural-network and ED states is evaluated using different Monte Carlo sample sizes at each filling factor. ${\rm 1e7}$ samples are used for $\nu=2/3$ and $5/7$, and more than ${\rm 2e5}$ samples are used for the other filling factors due to their million-dimensional Hilbert spaces.


\begin{table*}[htbp]
\centering
\caption{\textbf{Abelian results.}  $S=N/\nu-2Q$ is the topological  shift. Energy is displayed in the convention of \(E=T+\kappa V\). $d$ is the degree of the $SO(3)$ Gauss quadrature which integrates angular momentum exactly to $L=d$. $n_{\rm LL}$ is the number of Landau levels kept in ED. 
The trial state $\Psi_{\rm T}$
is the Laughlin state, or the projected composite-fermion state CF$(n,2p)$ at $\nu=n/(2pn+1)$. NA  marks system whose Hilbert space is too large for ED.
}
\label{tab:jain}
\footnotesize
\setlength{\extrarowheight}{2pt}
\setlength{\tabcolsep}{4pt}
\begin{tabular}{|c|l|c|c|c|c|c|c|c|c|c|c|c|}\hline
$\nu$ & $\Psi_{\rm T}$ & $N$ & $2Q$ & $S$ & $\kappa$ & $d$ & $n_{\rm LL}$ & $E_{\rm ED}/\hbar\omega_c$ & $E_{\rm Net}/\hbar\omega_c$ & $L^2/\hbar^2$
 & $|\langle\Psi_{\rm T}|\Psi_{\rm Net}\rangle|^2$ & $|\langle\Psi_{\rm ED}|\Psi_{\rm Net}\rangle|^2$ \\\hline
$1$ & IQHE & 12 & 11 & $1$ & $1$ & 5 & 1 & 28.764421 & 28.647049(2) & 0.000047(15) & 0.96732(3) & 0.96732(3) \\\hline
\multirow{6}{*}{$1/3$} & \multirow{6}{*}{Laughlin} & 8 & 21 & \multirow{6}{*}{$3$} & \multirow{4}{*}{$1$} & 5 & 2 & 10.356229 & 10.354635(4) & 0.00235(16) & 0.97873(2) & 0.999498(8) \\\cline{3-4}\cline{7-13}
 &  & 10 & 27 &  &  & \multirow{5}{*}{11} & \multirow{2}{*}{1} & 14.279867 & 14.269802(11) & 0.0404(8) & 0.96791(3) & 0.991680(15) \\\cline{3-4}\cline{9-13}
 &  & 12 & 33 &  &  &  &  & 18.579517 & 18.567017(10) & 0.0429(10) & 0.96053(3) & \multirow{4}{*}{NA} \\\cline{3-4}\cline{8-12}
 &  & \multirow{3}{*}{20} & \multirow{3}{*}{57} &  &  &  & \multirow{3}{*}{NA} & \multirow{3}{*}{NA} & 39.020367(10) & 0.0763(19) & 0.93047(5) &  \\\cline{6-6}\cline{10-12}
 &  &  &  &  & $5$ &  &  &  & 154.749833(14) & 0.074(3) & 0.6553(4) &  \\\cline{6-6}\cline{10-12}
 &  &  &  &  & $10$ &  &  &  & 298.878633(18) & 0.062(3) & 0.3621(16) &  \\\hline
$2/3$ & PH[Laughlin] & 12 & 18 & $0$ & $1$ & 5 & 1 & 23.653465 & 23.541244(7) & 0.0093(4) & 0.94279(5) & 0.93889(15) \\\hline
\multirow{3}{*}{$1/5$} & \multirow{3}{*}{Laughlin} & 7 & 30 & \multirow{3}{*}{$5$} & \multirow{3}{*}{$1$} & 5 & 2 & 7.349619 & 7.349492(5) & 0.0136(5) & 0.94888(7) & 0.999750(4) \\\cline{3-4}\cline{7-13}
 &  & 10 & 45 &  &  & \multirow{2}{*}{11} & 1 & 12.086418 & 12.083423(8) & 0.0871(18) & 0.81291(16) & \multirow{2}{*}{NA} \\\cline{3-4}\cline{8-12}
 &  & 12 & 55 &  &  &  & NA & NA & 15.616917(14) & 0.265(3) & 0.67704(18) &  \\\hline
$2/5$ & CF$(2,2)$ & 10 & 21 & $4$ & $1$ & 5 & 1 & 15.735838 & 15.683879(6) & 0.0068(3) & 0.96128(3) & 0.96806(5) \\\hline
$3/5$ & PH[CF$(2,2)$] & 12 & 21 & $-1$ & $1$ & 5 & 1 & 22.235069 & 22.141332(9) & 0.0161(5) & 0.94391(7) & 0.94285(8) \\\hline
$4/5$ & PH[Laughlin] & 12 & 15 & $0$ & $1$ & 5 & 2 & 25.490721 & 25.441241(4) & 0.00087(12) & 0.88168(13) & 0.98812(6) \\\hline
\multirow{10}{*}{$1/7$} & \multirow{10}{*}{Laughlin} & 6 & 35 & \multirow{10}{*}{$7$} & $1$ & 5 & 3 & 5.456936 & 5.457081(5) & 0.0733(18) & 0.7336(4) & 0.998737(13) \\\cline{3-4}\cline{6-13}
 &  & \multirow{8}{*}{10} & \multirow{8}{*}{63} &  & $0.5$ & \multirow{9}{*}{11} & \multirow{9}{*}{NA} & \multirow{9}{*}{NA} & 7.972875(5) & 0.362(4) & 0.6375(4) & \multirow{9}{*}{NA} \\\cline{6-6}\cline{10-12}
 &  &  &  &  & $1$ &  &  &  & 10.945274(9) & 0.889(7) & 0.5870(6) &  \\\cline{6-6}\cline{10-12}
 &  &  &  &  & $1.5$ &  &  &  & 13.916444(8) & 0.255(4) & 0.6006(6) &  \\\cline{6-6}\cline{10-12}
 &  &  &  &  & $2$ &  &  &  & 16.887127(9) & 0.174(3) & 0.5976(6) &  \\\cline{6-6}\cline{10-12}
 &  &  &  &  & $3$ &  &  &  & 22.826388(9) & 0.182(3) & 0.5968(7) &  \\\cline{6-6}\cline{10-12}
 &  &  &  &  & $4$ &  &  &  & 28.763363(10) & 0.158(2) & 0.5793(11) &  \\\cline{6-6}\cline{10-12}
 &  &  &  &  & $5$ &  &  &  & 34.697920(9) & 0.132(2) & 0.5674(8) &  \\\cline{6-6}\cline{10-12}
 &  &  &  &  & $10$ &  &  &  & 64.341898(12) & 0.159(3) & 0.4879(14) &  \\\cline{3-4}\cline{6-6}\cline{10-12}
 &  & 12 & 77 &  & $1$ &  &  &  & 14.073872(6) & 0.241(6) & 0.2397(8) &  \\\hline
$2/7$ & CF$(-2,4)$ & 8 & 26 & $2$ & $1$ & 5 & 2 & 9.682728 & 9.681960(6) & 0.0152(5) & 0.989641(9) & 0.999586(6) \\\hline
$3/7$ & CF$(3,2)$ & 15 & 30 & $5$ & $1$ & 5 & 1 & 29.444433 & 29.347477(16) & 0.1073(17) & 0.93298(6) & 0.9379(3) \\\hline
\multirow{2}{*}{$4/7$} & \multirow{2}{*}{CF$(-4,2)$} & 12 & 23 & \multirow{2}{*}{$-2$} & \multirow{2}{*}{$1$} & \multirow{2}{*}{5} & \multirow{2}{*}{1} & 21.439976 & 21.359213(13) & 0.0471(9) & 0.94349(5) & 0.94989(10) \\\cline{3-4}\cline{9-13}
 &  & 16 & 30 &  &  &  &  & 33.306927 & 33.189021(20) & 0.1378(16) & 0.92107(8) & 0.9219(4) \\\hline
$5/7$ & PH[CF$(-2,4)$] & 14 & 19 & $3/5$ & $1$ & 5 & 1 & 31.193481 & 31.060305(11) & 0.0335(7) & 0.94146(5) & 0.93422(6) \\\hline
$6/7$ & PH[Laughlin] & 12 & 14 & $0$ & $1$ & 5 & 2 & 26.219125 & 26.166584(3) & 0.00041(8) & 0.93957(4) & 0.98788(6) \\\hline
\multirow{2}{*}{$1/9$} & \multirow{2}{*}{Laughlin} & 10 & 81 & \multirow{2}{*}{$9$} & \multirow{2}{*}{$1$} & \multirow{2}{*}{11} & \multirow{2}{*}{NA} & \multirow{2}{*}{NA} & 10.221272(7) & 0.662(9) & 0.5317(14) & \multirow{2}{*}{NA} \\\cline{3-4}\cline{10-12}
 &  & 12 & 99 &  &  &  &  &  & 13.09213(2) & 2.39(5) & 0.123(2) &  \\\hline
\multirow{2}{*}{$1/2$} & \multirow{2}{*}{CFL} & 9 & 16 & \multirow{2}{*}{$2$} & \multirow{2}{*}{$1$} & 5 & 2 & 14.272003 & 14.261507(4) & 0.00238(16) & 0.96260(3) & 0.997406(17) \\\cline{3-4}\cline{7-13}
 &  & 25 & 48 &  &  & 11 & NA & NA & 64.17346(3) & 0.999(8) & 0.86307(13) & NA \\\hline
\end{tabular}
\end{table*}


\begin{table*}[htbp]
\centering
\caption{\textbf{States at $\nu=3/2$ as a function of $\kappa$.} Each state is computed at its own flux: Pf,
a-Pf and PH-Pf. ED keeps two
Landau levels, and every ED ground state listed has $L=0$.}
\label{tab:nu32}
\footnotesize
\setlength{\extrarowheight}{2pt}
\setlength{\tabcolsep}{4pt}
\begin{tabular}{|l|c|c|c|c|c|c|c|c|c|}\hline
State & $N$ & $2Q$ & $S$ & $d$ & $\kappa$ & $E_{\rm ED}/\hbar\omega_c$ & $E_{\rm Net}/\hbar\omega_c$ & $L^2/\hbar^2$
 & $|\langle\Psi_{\rm T}|\Psi_{\rm Net}\rangle|^2$ \\\hline
\multirow{9}{*}{Pf} & 8 & 3 & \multirow{9}{*}{$7/3$} & 5 & $1$ & 28.630939 & 28.511237(4) & 0.000009(4) & 0.981723(17) \\\cline{2-3}\cline{5-10}
 & \multirow{8}{*}{20} & \multirow{8}{*}{11} &  & \multirow{8}{*}{9} & $0.5$ & 53.788354 & 53.705710(16) & 0.0204(6) & 0.6605(3) \\\cline{6-10}
 &  &  &  &  & $1$ & 88.115588 & 87.80892(2) & 0.0651(10) & 0.5740(4) \\\cline{6-10}
 &  &  &  &  & $1.5$ & 122.437984 & 121.792593(18) & 0.0513(9) & 0.4983(4) \\\cline{6-10}
 &  &  &  &  & $2$ & 156.756719 & 155.67811(3) & 0.0473(10) & 0.4315(5) \\\cline{6-10}
 &  &  &  &  & $3$ & 225.386284 & 223.20721(3) & 0.0360(9) & 0.3276(6) \\\cline{6-10}
 &  &  &  &  & $4$ & 294.008564 & 290.478093(17) & 0.0094(4) & 0.2488(15) \\\cline{6-10}
 &  &  &  &  & $5$ & 362.625973 & 357.54635(4) & 0.0532(11) & 0.1942(11) \\\cline{6-10}
 &  &  &  &  & $10$ & 705.676590 & 690.80811(6) & 0.0829(15) & 0.0588(18) \\\hline
\multirow{9}{*}{a-Pf} & 12 & 7 & \multirow{9}{*}{$1$} & 5 & $1$ & 39.837466 & 39.672382(5) & 0.00021(5) & 0.5956(4) \\\cline{2-3}\cline{5-10}
 & \multirow{8}{*}{18} & \multirow{8}{*}{11} &  & \multirow{8}{*}{9} & $0.5$ & 43.504689 & 43.437788(10) & 0.0276(8) & 0.7251(5) \\\cline{6-10}
 &  &  &  &  & $1$ & 70.908859 & 70.65978(2) & 0.0470(8) & 0.6873(3) \\\cline{6-10}
 &  &  &  &  & $1.5$ & 98.306182 & 97.77990(2) & 0.0502(10) & 0.6446(4) \\\cline{6-10}
 &  &  &  &  & $2$ & 125.698461 & 124.81453(3) & 0.0532(12) & 0.5907(4) \\\cline{6-10}
 &  &  &  &  & $3$ & 180.472402 & 178.67285(2) & 0.0282(8) & 0.4780(6) \\\cline{6-10}
 &  &  &  &  & $4$ & 235.236765 & 232.30705(4) & 0.0771(14) & 0.3774(15) \\\cline{6-10}
 &  &  &  &  & $5$ & 289.994815 & 285.76094(5) & 0.0709(14) & 0.3017(18) \\\cline{6-10}
 &  &  &  &  & $10$ & 563.736972 & 551.25581(3) & 0.0228(8) & 0.105(3) \\\hline
\multirow{10}{*}{PH-Pf} & 10 & 5 & \multirow{10}{*}{$5/3$} & 5 & $1$ & 33.418010 & 33.283976(4) & 0.000020(13) & 0.97345(2) \\\cline{2-3}\cline{5-10}
 & 16 & 9 &  & 9 & $1$ & 62.484994 & 62.240077(11) & 0.0044(3) & 0.000060(4) \\\cline{2-3}\cline{5-10}
 & \multirow{8}{*}{22} & \multirow{8}{*}{13} &  & \multirow{8}{*}{11} & $0.5$ & 58.874348 & 58.787846(20) & 0.231(2) & 0.000079(5) \\\cline{6-10}
 &  &  &  &  & $1$ & 97.507984 & 97.18576(2) & 0.332(3) & 0.001031(17) \\\cline{6-10}
 &  &  &  &  & $1.5$ & 136.134429 & 135.45639(3) & 0.349(3) & 0.00156(2) \\\cline{6-10}
 &  &  &  &  & $2$ & 174.755495 & 173.61865(3) & 0.237(3) & 0.00271(3) \\\cline{6-10}
 &  &  &  &  & $3$ & 251.986109 & 249.68181(5) & 0.539(4) & 0.00364(4) \\\cline{6-10}
 &  &  &  &  & $4$ & 329.206167 & 325.45326(4) & 0.213(2) & 0.00337(4) \\\cline{6-10}
 &  &  &  &  & $5$ & 406.419186 & 401.00000(5) & 0.200(3) & 0.00239(4) \\\cline{6-10}
 &  &  &  &  & $10$ & 792.431482 & 776.49523(8) & 0.218(3) & 0.00056(10) \\\hline
\end{tabular}
\end{table*}

\begin{table*}[htbp]
\centering
\caption{\textbf{States at $\nu=4/3$ as a function of $\kappa$.}  Different topological shifts are used for different states. $S_a$  denotes the active-shell shift
$S_a=N_a/\nu_a-2(Q+1)$, with $N_a=N-(2Q+1)$ electrons in the 1LL and
orbital degeneracy $2Q+3$. 
Every trial state includes the filled lowest Landau level. 
ED keeps two Landau levels, and every ED ground state listed has $L=0$. NA marks systems whose Hilbert space is too large for ED. }
\label{tab:nu43}
\footnotesize
\setlength{\extrarowheight}{2pt}
\setlength{\tabcolsep}{4pt}
\begin{tabular}{|l|c|c|c|c|c|c|c|c|c|l|c|}\hline
$\nu$ & $N$ & $2Q$ & $S$ & $S_a$ & $d$ & $\kappa$ & $E_{\rm ED}/\hbar\omega_c$ & $E_{\rm Net}/\hbar\omega_c$ & $L^2/\hbar^2$ & $\Psi_{\rm T}$
 & $|\langle\Psi_{\rm T}|\Psi_{\rm Net}\rangle|^2$ \\\hline
\multirow{51}{*}{4/3} & \multirow{14}{*}{18} & \multirow{14}{*}{13} & \multirow{18}{*}{$1/2$} & \multirow{18}{*}{$-3$} & \multirow{14}{*}{9} & \multirow{2}{*}{$0.5$} & \multirow{2}{*}{38.780522} & \multirow{2}{*}{38.728771(8)} & \multirow{2}{*}{0.0043(3)} & a-RR4 & 0.87792(13) \\\cline{11-12}
 &  &  &  &  &  &  &  &  &  & $\bar{2}^41^5$ & 0.93363(7) \\\cline{7-12}
 &  &  &  &  &  & \multirow{2}{*}{$1$} & \multirow{2}{*}{63.924759} & \multirow{2}{*}{63.730526(6)} & \multirow{2}{*}{0.00122(16)} & a-RR4 & 0.82351(17) \\\cline{11-12}
 &  &  &  &  &  &  &  &  &  & $\bar{2}^41^5$ & 0.88034(11) \\\cline{7-12}
 &  &  &  &  &  & \multirow{2}{*}{$1.5$} & \multirow{2}{*}{89.053451} & \multirow{2}{*}{88.641495(13)} & \multirow{2}{*}{0.0081(4)} & a-RR4 & 0.7584(2) \\\cline{11-12}
 &  &  &  &  &  &  &  &  &  & $\bar{2}^41^5$ & 0.81465(18) \\\cline{7-12}
 &  &  &  &  &  & \multirow{2}{*}{$2$} & \multirow{2}{*}{114.170312} & \multirow{2}{*}{113.475471(13)} & \multirow{2}{*}{0.0036(3)} & a-RR4 & 0.6893(4) \\\cline{11-12}
 &  &  &  &  &  &  &  &  &  & $\bar{2}^41^5$ & 0.7442(3) \\\cline{7-12}
 &  &  &  &  &  & \multirow{2}{*}{$2.5$} & \multirow{2}{*}{139.277982} & \multirow{2}{*}{138.243668(11)} & \multirow{2}{*}{0.00175(20)} & a-RR4 & 0.6211(6) \\\cline{11-12}
 &  &  &  &  &  &  &  &  &  & $\bar{2}^41^5$ & 0.6734(5) \\\cline{7-12}
 &  &  &  &  &  & \multirow{2}{*}{$3$} & \multirow{2}{*}{164.378383} & \multirow{2}{*}{162.955063(17)} & \multirow{2}{*}{0.0054(3)} & a-RR4 & 0.5582(6) \\\cline{11-12}
 &  &  &  &  &  &  &  &  &  & $\bar{2}^41^5$ & 0.6075(6) \\\cline{7-12}
 &  &  &  &  &  & \multirow{2}{*}{$5$} & \multirow{2}{*}{264.731213} & \multirow{2}{*}{261.347382(10)} & \multirow{2}{*}{0.00122(18)} & a-RR4 & 0.355(3) \\\cline{11-12}
 &  &  &  &  &  &  &  &  &  & $\bar{2}^41^5$ & 0.392(2) \\\cline{2-3}\cline{6-12}
 & \multirow{2}{*}{26} & \multirow{2}{*}{19} &  &  & \multirow{4}{*}{11} & \multirow{4}{*}{$1$} & \multirow{4}{*}{NA} & \multirow{2}{*}{110.074692(13)} & \multirow{2}{*}{0.0684(13)} & a-RR4 & 0.5823(3) \\\cline{11-12}
 &  &  &  &  &  &  &  &  &  & $\bar{2}^41^5$ & 0.79661(18) \\\cline{2-3}\cline{9-12}
 & \multirow{2}{*}{34} & \multirow{2}{*}{25} &  &  &  &  &  & \multirow{2}{*}{164.15492(4)} & \multirow{2}{*}{2.782(8)} & a-RR4 & 0.6946(4) \\\cline{11-12}
 &  &  &  &  &  &  &  &  &  & $\bar{2}^41^5$ & 0.7856(2) \\\cline{2-12}
 & \multirow{3}{*}{12} & \multirow{3}{*}{7} & \multirow{33}{*}{$2$} & \multirow{33}{*}{$3$} & \multirow{3}{*}{5} & \multirow{3}{*}{$1$} & \multirow{3}{*}{39.837466} & \multirow{3}{*}{39.672382(5)} & \multirow{3}{*}{0.00021(5)} & CF & 0.1821(3) \\\cline{11-12}
 &  &  &  &  &  &  &  &  &  & $Z_1$ & 0.2221(3) \\\cline{11-12}
 &  &  &  &  &  &  &  &  &  & $Z_2$ & 0.4289(4) \\\cline{2-3}\cline{6-12}
 & \multirow{27}{*}{20} & \multirow{27}{*}{13} &  &  & \multirow{27}{*}{9} & \multirow{3}{*}{$0.5$} & \multirow{3}{*}{48.467155} & \multirow{3}{*}{48.394705(14)} & \multirow{3}{*}{0.291(3)} & CF & 0.1801(3) \\\cline{11-12}
 &  &  &  &  &  &  &  &  &  & $Z_1$ & 0.2245(4) \\\cline{11-12}
 &  &  &  &  &  &  &  &  &  & $Z_2$ & 0.4323(4) \\\cline{7-12}
 &  &  &  &  &  & \multirow{3}{*}{$1$} & \multirow{3}{*}{79.995654} & \multirow{3}{*}{79.724726(18)} & \multirow{3}{*}{0.311(3)} & CF & 0.1404(3) \\\cline{11-12}
 &  &  &  &  &  &  &  &  &  & $Z_1$ & 0.1792(3) \\\cline{11-12}
 &  &  &  &  &  &  &  &  &  & $Z_2$ & 0.3658(5) \\\cline{7-12}
 &  &  &  &  &  & \multirow{3}{*}{$1.5$} & \multirow{3}{*}{111.513033} & \multirow{3}{*}{110.94174(2)} & \multirow{3}{*}{0.363(3)} & CF & 0.1087(2) \\\cline{11-12}
 &  &  &  &  &  &  &  &  &  & $Z_1$ & 0.1421(3) \\\cline{11-12}
 &  &  &  &  &  &  &  &  &  & $Z_2$ & 0.3063(4) \\\cline{7-12}
 &  &  &  &  &  & \multirow{3}{*}{$2$} & \multirow{3}{*}{143.022201} & \multirow{3}{*}{142.06379(3)} & \multirow{3}{*}{0.245(3)} & CF & 0.0914(3) \\\cline{11-12}
 &  &  &  &  &  &  &  &  &  & $Z_1$ & 0.1209(4) \\\cline{11-12}
 &  &  &  &  &  &  &  &  &  & $Z_2$ & 0.2678(6) \\\cline{7-12}
 &  &  &  &  &  & \multirow{3}{*}{$2.5$} & \multirow{3}{*}{174.525126} & \multirow{3}{*}{173.104492(14)} & \multirow{3}{*}{0.0226(8)} & CF & 0.0769(3) \\\cline{11-12}
 &  &  &  &  &  &  &  &  &  & $Z_1$ & 0.1027(4) \\\cline{11-12}
 &  &  &  &  &  &  &  &  &  & $Z_2$ & 0.2330(6) \\\cline{7-12}
 &  &  &  &  &  & \multirow{3}{*}{$3$} & \multirow{3}{*}{206.023188} & \multirow{3}{*}{204.07966(3)} & \multirow{3}{*}{0.1390(19)} & CF & 0.0673(3) \\\cline{11-12}
 &  &  &  &  &  &  &  &  &  & $Z_1$ & 0.0902(3) \\\cline{11-12}
 &  &  &  &  &  &  &  &  &  & $Z_2$ & 0.2065(6) \\\cline{7-12}
 &  &  &  &  &  & \multirow{3}{*}{$4$} & \multirow{3}{*}{269.008436} & \multirow{3}{*}{265.85515(4)} & \multirow{3}{*}{0.1203(18)} & CF & 0.0498(4) \\\cline{11-12}
 &  &  &  &  &  &  &  &  &  & $Z_1$ & 0.0676(5) \\\cline{11-12}
 &  &  &  &  &  &  &  &  &  & $Z_2$ & 0.1592(9) \\\cline{7-12}
 &  &  &  &  &  & \multirow{3}{*}{$5$} & \multirow{3}{*}{331.983244} & \multirow{3}{*}{327.43928(4)} & \multirow{3}{*}{0.0719(14)} & CF & 0.0378(4) \\\cline{11-12}
 &  &  &  &  &  &  &  &  &  & $Z_1$ & 0.0517(6) \\\cline{11-12}
 &  &  &  &  &  &  &  &  &  & $Z_2$ & 0.1229(14) \\\cline{7-12}
 &  &  &  &  &  & \multirow{3}{*}{$10$} & \multirow{3}{*}{646.778612} & \multirow{3}{*}{633.48901(5)} & \multirow{3}{*}{0.0622(17)} & CF & 0.0126(5) \\\cline{11-12}
 &  &  &  &  &  &  &  &  &  & $Z_1$ & 0.0174(6) \\\cline{11-12}
 &  &  &  &  &  &  &  &  &  & $Z_2$ & 0.0419(16) \\\cline{2-3}\cline{6-12}
 & \multirow{3}{*}{28} & \multirow{3}{*}{19} &  &  & \multirow{3}{*}{11} & \multirow{3}{*}{$1$} & \multirow{3}{*}{NA} & \multirow{3}{*}{128.64194(4)} & \multirow{3}{*}{1.920(7)} & CF & 0.1171(2) \\\cline{11-12}
 &  &  &  &  &  &  &  &  &  & $Z_1$ & 0.1492(3) \\\cline{11-12}
 &  &  &  &  &  &  &  &  &  & $Z_2$ & 0.3230(4) \\\hline
\end{tabular}
\end{table*}

\begin{table*}[htbp]
\centering
\caption{\textbf{States at $\nu=7/5$ as a function of $\kappa$.} Different topological shifts are used for different states. 
}
\label{tab:nu75}
\footnotesize
\setlength{\extrarowheight}{2pt}
\setlength{\tabcolsep}{4pt}
\begin{tabular}{|l|c|c|c|c|c|c|c|c|c|l|c|}\hline
$\nu$ & $N$ & $2Q$ & $S$ & $S_a$ & $d$ & $\kappa$ & $E_{\rm ED}/\hbar\omega_c$ & $E_{\rm Net}/\hbar\omega_c$ & $L^2/\hbar^2$ & $\Psi_{\rm T}$
 & $|\langle\Psi_{\rm T}|\Psi_{\rm Net}\rangle|^2$ \\\hline
\multirow{23}{*}{7/5} & \multirow{2}{*}{15} & \multirow{2}{*}{10} & \multirow{14}{*}{$5/7$} & \multirow{14}{*}{$-2$} & \multirow{2}{*}{5} & \multirow{2}{*}{$1$} & \multirow{2}{*}{51.152898} & \multirow{2}{*}{50.976577(6)} & \multirow{2}{*}{0.00068(8)} & a-RR3 & 0.91731(7) \\\cline{11-12}
 &  &  &  &  &  &  &  &  &  & $\bar{2}^31^4$ & 0.94701(5) \\\cline{2-3}\cline{6-12}
 & \multirow{12}{*}{22} & \multirow{12}{*}{15} &  &  & \multirow{12}{*}{9} & \multirow{2}{*}{$0.5$} & \multirow{2}{*}{53.728305} & \multirow{2}{*}{53.653792(9)} & \multirow{2}{*}{0.245(2)} & a-RR3 & 0.7309(2) \\\cline{11-12}
 &  &  &  &  &  &  &  &  &  & $\bar{2}^31^4$ & 0.8722(3) \\\cline{7-12}
 &  &  &  &  &  & \multirow{2}{*}{$1$} & \multirow{2}{*}{89.637884} & \multirow{2}{*}{89.358791(14)} & \multirow{2}{*}{0.1095(16)} & a-RR3 & 0.5833(3) \\\cline{11-12}
 &  &  &  &  &  &  &  &  &  & $\bar{2}^31^4$ & 0.7429(4) \\\cline{7-12}
 &  &  &  &  &  & \multirow{2}{*}{$1.5$} & \multirow{2}{*}{125.533613} & \multirow{2}{*}{124.943093(19)} & \multirow{2}{*}{0.1300(18)} & a-RR3 & 0.4378(4) \\\cline{11-12}
 &  &  &  &  &  &  &  &  &  & $\bar{2}^31^4$ & 0.6013(5) \\\cline{7-12}
 &  &  &  &  &  & \multirow{2}{*}{$2$} & \multirow{2}{*}{161.418856} & \multirow{2}{*}{160.424992(18)} & \multirow{2}{*}{0.0609(12)} & a-RR3 & 0.3157(4) \\\cline{11-12}
 &  &  &  &  &  &  &  &  &  & $\bar{2}^31^4$ & 0.4704(7) \\\cline{7-12}
 &  &  &  &  &  & \multirow{2}{*}{$2.5$} & \multirow{2}{*}{197.295994} & \multirow{2}{*}{195.820542(18)} & \multirow{2}{*}{0.0628(13)} & a-RR3 & 0.2349(5) \\\cline{11-12}
 &  &  &  &  &  &  &  &  &  & $\bar{2}^31^4$ & 0.3722(9) \\\cline{7-12}
 &  &  &  &  &  & \multirow{2}{*}{$3$} & \multirow{2}{*}{233.166754} & \multirow{2}{*}{231.141096(20)} & \multirow{2}{*}{0.0453(9)} & a-RR3 & 0.1780(6) \\\cline{11-12}
 &  &  &  &  &  &  &  &  &  & $\bar{2}^31^4$ & 0.2977(11) \\\cline{2-12}
 & 9 & 4 & \multirow{9}{*}{$17/7$} & \multirow{9}{*}{$4$} & 5 & $1$ & 30.752167 & 30.681054(3) & 0.000016(7) & \multirow{9}{*}{CF} & 0.979031(19) \\\cline{2-3}\cline{6-10}\cline{12-12}
 & \multirow{7}{*}{16} & \multirow{7}{*}{9} &  &  & \multirow{8}{*}{9} & $0.5$ & 38.913814 & 38.847270(8) & 0.0043(2) &  & 0.000194(7) \\\cline{7-10}\cline{12-12}
 &  &  &  &  &  & $1$ & 62.484994 & 62.240077(11) & 0.0044(3) &  & 0.000191(8) \\\cline{7-10}\cline{12-12}
 &  &  &  &  &  & $1.5$ & 86.048997 & 85.536705(13) & 0.0070(3) &  & 0.000123(5) \\\cline{7-10}\cline{12-12}
 &  &  &  &  &  & $2$ & 109.607460 & 108.753990(14) & 0.0059(3) &  & 0.000027(3) \\\cline{7-10}\cline{12-12}
 &  &  &  &  &  & $2.5$ & 133.161621 & 131.904022(16) & 0.0109(4) &  & 0.0000021(8) \\\cline{7-10}\cline{12-12}
 &  &  &  &  &  & $3$ & 156.712407 & 154.996078(17) & 0.0076(3) &  & 0.0000028(8) \\\cline{7-10}\cline{12-12}
 &  &  &  &  &  & $20$ & 956.966139 & 925.96423(5) & 0.0134(7) &  & 0.00035(3) \\\cline{2-3}\cline{7-10}\cline{12-12}
 & 23 & 14 &  &  &  & $1$ & 102.436878 & 102.10519(3) & 1.101(5) &  & 0.00530(4) \\\hline
\end{tabular}
\end{table*}

\section{Analytical reference states}

We consider several analytical reference states, including Pfaffian,
Read--Rezayi, parton, and composite-fermion states.
On the sphere, we define
$\omega_{ij}=u_iv_j-u_jv_i$, where $(u_i,v_i)$ are spinor
coordinates. The Pfaffian and PH-Pfaffian states are
constructed as
\begin{equation}
  \Phi_{\rm Pf}
  \propto \operatorname{Pf}\left(\frac{1}{\omega_{ij}}\right)\left(\prod_{i<j}\omega_{ij}\right)^2,
  \qquad
  \Phi_{\rm PH\text{-}Pf}
  \propto \mathcal P_{\rm LLL}
  \left[
    \operatorname{Pf}\left(\frac{1}{\omega_{ij}^*}\right)\left(\prod_{i<j}\omega_{ij}\right)^2
  \right],
\end{equation}
where $\mathcal P_{\rm LLL}$ denotes projection onto the LLL.
The anti-Pfaffian is obtained by particle–hole conjugation
of the Pfaffian.

The Read--Rezayi states are constructed by symmetrizing over
$k$ equal-sized clusters,
\begin{equation}
  \Phi_{{\rm RRk}}
  \propto
  \left(\prod_{i<j}\omega_{ij}\right)\,\mathcal S
  \left[
    \prod_{a=1}^{k}
    \prod_{i<j\in G_a}\omega_{ij}^2
  \right],
\end{equation}
where $\mathcal S$ symmetrizes over electron partitions
into clusters $G_a$.

Parton states are constructed by multiplying integer quantum Hall
wavefunctions and projecting onto the LLL,
\begin{equation}
  \Phi_{\rm parton}
  \propto \mathcal P_{\rm LLL}
  \left[\prod_a\Phi_{n_a}\right],
\end{equation}
where $\Phi_n$ denotes $n$ filled Landau levels,
with $\Phi_{-n}=\Phi_n^*$.

Composite-fermion states are constructed by attaching
$2p$ fluxes to integer quantum Hall states,
\begin{equation}
  \Phi_{\rm CF}
  \propto \mathcal P_{\rm LLL}
  \left[\Phi_n \left(\prod_{i<j}\omega_{ij}\right)^{2p}\right].
\end{equation}


To construct target wavefunctions explicitly including LLL and 1LL, we first expand the corresponding analytical
states in the LLL basis.
Let $N$ denote the total electron number and
$N_a=N-(2Q+1)$ the number of electrons in the active 1LL.
The expansion is
\begin{equation}
  |\Phi_{\rm LLL}\rangle
  =
  \sum_\alpha c_\alpha^{\rm ref}
  |m_1^{\alpha},\ldots,m_{N_a}^{\alpha}\rangle_{\rm LLL}.
  \label{eq:si_lll_fock_expansion}
\end{equation}
We then transplant the state to the 1LL by preserving its
expansion coefficients while promoting each LLL basis state
to its corresponding 1LL counterpart,
\begin{equation}
  |m_1,\ldots,m_{N_a}\rangle_{\rm LLL}
  \longrightarrow
  |m_1,\ldots,m_{N_a}\rangle_{\rm 1LL}.
  \label{eq:si_ll_transplant}
\end{equation}

For example, for the Pfaffian wavefunction at $\nu=3/2$, the
$N_a$-electron LLL Pfaffian has
$2Q_{\rm eff}=2N_a-3$, while the physical system has
$2Q=2N_a-5$. Hence
\begin{equation}
  Q_{\rm eff}=Q+1,
\end{equation}
so that the auxiliary LLL and the physical 1LL have the same single-particle
angular momentum and the same set of magnetic quantum numbers $m$.
Including the filled physical LLL, the transplanted reference state is
\begin{equation}
  \Phi_{3/2}(X)
  =
  \sum_\alpha c_\alpha^{\rm ref}
  \det\!\left[
    \{\phi^{\rm LLL}_{Q,Q,m}\}_{m=-Q}^{Q},
    \{\phi^{\rm 1LL}_{Q,Q+1,m}\}_{m\in\alpha}
  \right].
  \label{eq:si_mr_sll_reference}
\end{equation}
The determinants include all $N$ electrons and are fully
antisymmetric. The resulting state can be evaluated directly in
real space and compared with the network using the squared-overlap
estimator.

\section{Pair correlation and conditional density}
For a rotation-invariant state on the sphere, the pair correlation
function $g(\theta)$ depends only on the geodesic angle
$\theta_{ij}=\arccos(\mathbf n_i\cdot\mathbf n_j)$ between
the unit-sphere directions $\mathbf n_i=\mathbf r_i/R$:
\begin{equation}
  g(\theta)
  =
  \frac{2}{N^2}
  \left\langle
    \sum_{i\ne j}
    \delta(\cos\theta-\mathbf n_i\cdot\mathbf n_j)
  \right\rangle,
  \label{eq:si_pair_correlation}
\end{equation}
normalized so that
$\frac12\int_0^\pi g(\theta)\sin\theta\,d\theta=(N-1)/N$.

To visualize crystalline correlations in real space, we also evaluate the
two-electron conditional density $\rho_{\rm cond}(\mathbf n\mid\mathbf n_a,\mathbf n_b)$,
defined as the density of the remaining $N-2$ electrons at unit-sphere
direction $\mathbf n=\mathbf r/R$ when two electrons are pinned at fixed
directions $\mathbf n_a$ and $\mathbf n_b$:
\begin{equation}
  \rho_{\rm cond}(\mathbf n\mid\mathbf n_a,\mathbf n_b)
  =
  \frac{1}{R^2}
  \frac{\left\langle
    \sum_{i\ne j}
    \delta(\mathbf n_i-\mathbf n_a)\,
    \delta(\mathbf n_j-\mathbf n_b)
    \sum_{k\ne i,j}\delta(\mathbf n-\mathbf n_k)
  \right\rangle}
  {\left\langle
    \sum_{i\ne j}
    \delta(\mathbf n_i-\mathbf n_a)\,
    \delta(\mathbf n_j-\mathbf n_b)
  \right\rangle},
  \label{eq:si_conditional_density}
\end{equation}
where $\delta(\mathbf n-\mathbf n')$ denotes the Dirac delta function with
respect to the unit-sphere solid angle $d\Omega$, so that
$\int\rho_{\rm cond}(\mathbf n\mid\mathbf n_a,\mathbf n_b)\,R^2 d\Omega=N-2$.

\section{Topological property}

We compute the squared overlap between the network and the analytical
reference states using wavefunction ratios. With
$X\sim |\Psi(X)|^2$ and
$f(X)=\Phi(X)/\Psi(X)$, the normalized squared overlap is
\begin{equation}
  \mathcal O^2
  =
  \frac{|\langle\Psi|\Phi\rangle|^2}
  {\langle\Psi|\Psi\rangle\langle\Phi|\Phi\rangle}
  =
  \frac{|\langle f\rangle|_{{|\Psi|^2}}^2}
  {\langle |f|^2\rangle_{{|\Psi|^2}}}.
  \label{eq:si_realspace_overlap}
\end{equation}

For the entanglement-spectrum calculations, we evaluated the
amplitudes of the neural-network wavefunction in the
Slater-determinant basis $\{D_\alpha\}$ using a similar formula:
\begin{equation}
  \tilde c_\alpha^{\rm net}=
  \left\langle\frac{D_\alpha^*(X)}{\Psi^*(X)}\right\rangle_{|\Psi|^2},
  \qquad
  c_\alpha^{\rm net}=\frac{\tilde c_\alpha^{\rm net}}
  {(\sum_\beta|\tilde c_\beta^{\rm net}|^2)^{1/2}}.
  \label{eq:si_fock_tomography}
\end{equation}

For the particle entanglement spectrum, we partition the $N$
electrons into a subsystem $A$ containing $N_A$ particles and its
complement $B$. The normalized reduced density matrix was obtained
by tracing out subsystem $B$,
\begin{equation}
  \rho_A^{\mathrm{PES}}
  = \operatorname{Tr}_B
  \left(
    |\Psi\rangle
    \langle\Psi|
  \right).
  \label{eq:si_pes}
\end{equation}
The partial trace was evaluated using the calculated
Slater-determinant coefficients, with fermionic exchange signs
properly accounted for. We diagonalized $\rho_A^{\mathrm{PES}}$
separately in each subsystem $L_z^A$ sector to obtain the
entanglement energies $\xi_a=-\ln\lambda_a$.

For the orbital entanglement spectrum, after the filled-LLL
core was removed, the normalized valence state was arranged as
$|\Psi_{\rm val}\rangle=\sum_{\alpha\beta}
C_{\alpha\beta}|\alpha\rangle_A|\beta\rangle_B$.
Singular-value decomposition of each $(N_A,L_z^A)$ block gives
$\lambda_a=|s_a|^2$ and $\xi_a=-\ln\lambda_a$.

\begin{figure*}[!htbp]
    \centering
    \includegraphics[width=0.96\textwidth]{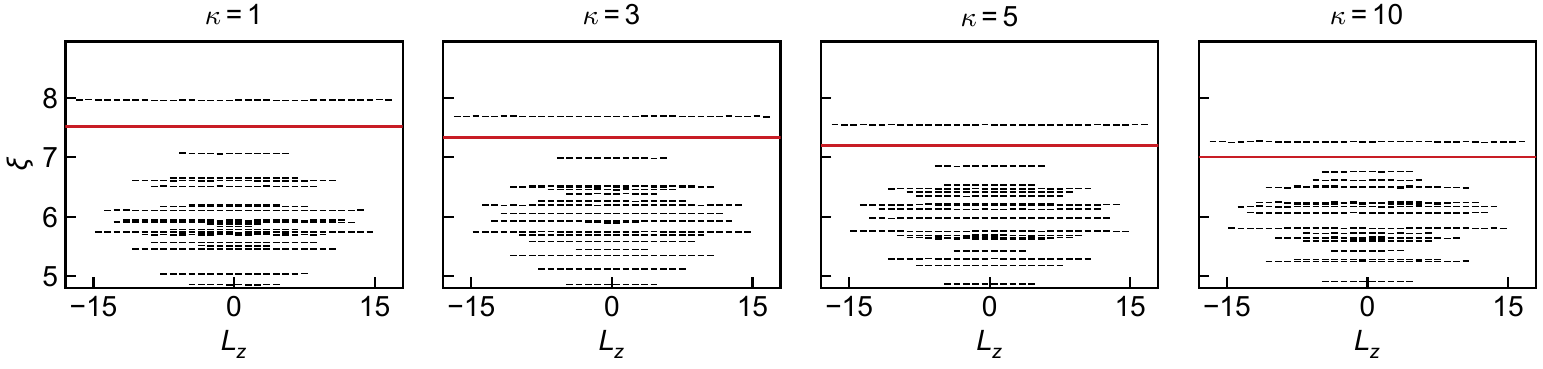}
    \caption{\textbf{Evolution of the Pfaffian particle entanglement spectrum with $\kappa$ at $\nu=3/2$.}
    Active-1LL $N_A=3$ PES at $(N,2Q)=(20,11)$ ($S_a=3$, $N_a=8$) evaluated from Eqs.~\eqref{eq:si_fock_tomography}--\eqref{eq:si_pes} at $\kappa=1,3,5,$ and $10$. The red line marks the entanglement gap separating the low-lying Moore--Read $(2,4)$-admissible quasihole levels from the higher levels.}
    \label{fig:si_pf_pes_kappa}
\end{figure*}

In Eq.~\eqref{eq:si_fock_tomography}, the Slater-determinant basis $\{D_\alpha\}$ spans only the subspace of a filled LLL and $N_a$ electrons in the active 1LL, so the PES is evaluated under a projection onto this two-shell subspace, whereas the real-space overlap in Eq.~\eqref{eq:si_realspace_overlap} is evaluated in the full continuum space including all Landau levels. At $\nu=3/2$, as $\kappa$ increases to $3, 5,$ and $10$, virtual excitations shift most of the wavefunction weight into higher Landau levels ($n\ge 2$), causing the full-space Pfaffian overlap to drop to nearly zero. By contrast, after projecting onto the active 1LL and normalizing via Eq.~\eqref{eq:si_fock_tomography}, the discarded higher-Landau-level leakage is removed, and the resulting $N_A=3$ PES still exhibits the Moore--Read $(2,4)$-admissible quasihole counting and a persistent entanglement gap up to $\kappa=10$ (Supplementary Fig.~\ref{fig:si_pf_pes_kappa}). This discrepancy demonstrates that projected diagnostics can mask the severe depletion of the target phase into higher Landau levels and thus overestimate the robustness of non-Abelian states against Landau-level mixing.

\section{Finite-size analysis}

\begin{figure}[H]
    \centering
    \includegraphics[width=0.4\textwidth]{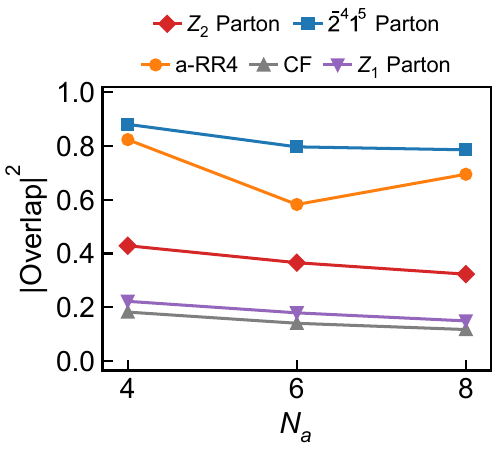}
    \caption{\textbf{Finite-size scaling of squared overlaps at $\nu=4/3$ and $\kappa=1$.}
    $N_a$  denotes the active-shell electron number in the 1LL.
    Results use $(N,2Q)=(18,13)$, $(26,19)$, and $(34,25)$ with $S_a=-3$ and $N_a=4,6,8$. For the $\bar{2}^41^5$ parton and a-RR4 states. For the $Z_2$ parton, $Z_1$ parton, and CF states, we use $(N,2Q)=(12,7)$, $(20,13)$, and $(28,19)$ with $S_a=3$.
    }
    \label{fig:si_nu43_na}
\end{figure}

For $\nu=4/3$ , Supplementary Fig.~\ref{fig:si_nu43_na} shows the overlap between the network and trial states for different numbers of active electrons $N_a$. The consistent ordering of the overlaps across $N_a$ supports the identification of the ground state as non-Abelian.

Moreover, the PES of the Laughlin sector, shown in Supplementary Fig.~\ref{fig:nu43_pes}, does not exhibit the expected counting of $(1,3)$-admissible configurations, providing further evidence against a Laughlin-type Abelian ground state.

\begin{figure}[htbp]
    \centering
    \includegraphics[
         width=0.4\columnwidth
     ]{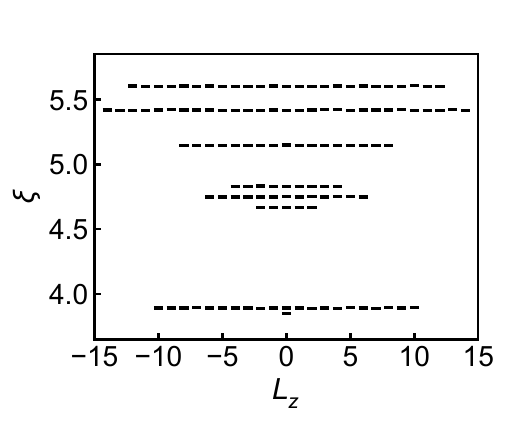}
    \caption{
        Particle entanglement spectrum of the
        $\nu=4/3$ Laughlin-sector network result.
        The calculation uses $(N,2Q)=(20,13)$ with
        $\kappa=1$ and $N_A=2$.
    \label{fig:nu43_pes}
    }

\end{figure}

\section{Neural-network architecture}

For each electron, the normalized spinor
$z_i=(u_i,v_i)^{\mathsf T}$ is mapped to the
unit-sphere coordinate
\begin{equation}
\mathbf n_i
=
z_i^\dagger\boldsymbol{\sigma}^{*}z_i
=
\left(
2\operatorname{Re}[u_i^*v_i],
-2\operatorname{Im}[u_i^*v_i],
|u_i|^2-|v_i|^2
\right),
\label{eq:spinor_feature}
\end{equation}
where
$\boldsymbol{\sigma}^{*}=(\sigma_x,-\sigma_y,\sigma_z)$
denotes the complex conjugate of the Pauli matrices.

These gauge-invariant features are processed by a
PsiFormer backbone~\cite{vonGlehn2023Psiformer}.
The input features are first embedded to a hidden
dimension $D=Hd_H$, where $H$ is the number of
attention heads and $d_H$ is the dimension of each
head. Each PsiFormer layer employs pre-layer
normalization, multi-head self-attention, and a
multilayer perceptron (MLP), with residual
connections:
\begin{equation}
\begin{aligned}
Y^{l+1}
&=X^l+\operatorname{MHA}(\operatorname{LN}(X^l)),\\
X^{l+1}
&=Y^{l+1}
+\operatorname{MLP}(\operatorname{LN}(Y^{l+1})).
\end{aligned}
\label{eq:psiformer_layer}
\end{equation}
Here $\operatorname{LN}$ denotes layer normalization.
The multi-head attention operation reads
\begin{equation}
\begin{aligned}
\operatorname{MHA}(X)
&=\operatorname{Concat}_{h=1}^{H}(A_hV_h)W_O+b_O,\\
A_h
&=\operatorname{softmax}
\left(\frac{Q_hK_h^{\mathsf T}}{\sqrt{d_H}}\right),\\
Q_h&=XW_{Q,h}+b_{Q,h},\\
K_h&=XW_{K,h}+b_{K,h},\\
V_h&=XW_{V,h}+b_{V,h}.
\end{aligned}
\label{eq:qkv_projection}
\end{equation}
The softmax is applied along the key dimension.

The final PsiFormer features are mapped to complex-valued functions
$f_{im}(\mathbf r_j;\mathbf r_{\neq j})$ through linear layers. 

\begin{equation}
\begin{aligned}
\phi_i(z_j;z_{\neq j})
&=
\sum_{m=-Q}^{Q}
f_{im}(\mathbf r_j;\mathbf r_{\neq j})
\mathcal Y_{Qm}(z_j),\\
\mathcal Y_{Qm}(z_j)
&=
\sqrt{C_{2Q}^{Q+m}}\,
u_j^{Q+m}v_j^{Q-m},\\
\psi^{\rm net}(z_1,\ldots,z_N)
&=
e^{J(\mathbf r_1,\ldots,\mathbf r_N)}
\det\!\left[\phi_i(z_j;z_{\neq j})\right].
\end{aligned}
\label{eq:si_net_ansatz}
\end{equation}

For fully spin-polarized states, the spherical
Jastrow exponent is
\begin{equation}
J
=
-\sum_{i<j}
\frac{c\,a^2}{a+\rho_{ij}},
\qquad
\rho_{ij}=|\mathbf n_i-\mathbf n_j|.
\label{eq:pade_jastrow}
\end{equation}
Here $\rho_{ij}$ is the chord distance on the unit
sphere. The parameter $a$ is optimized
together with the network parameters, while $c$
fixes the short-distance cusp. With the energy and length units specified above, the spin-polarized
Coulomb cusp fixes $c=\kappa R/(3\ell_B)=\kappa\sqrt{Q}/3$
.


The constructed wavefunction does not enforce a definite total angular momentum.
We construct rotational singlets using the \(L=0\) projector
\begin{equation}
\psi_{L=0}^{\rm net}(\mathbf r_1,\ldots,\mathbf r_N)
=
\int_{SO(3)} dM\,
\hat U_Q(M)\,
\psi^{\rm net}(\mathbf r_1,\ldots,\mathbf r_N),
\label{eq:L0_projection}
\end{equation}
where \(dM\) is the normalized Haar measure and \(\hat U_Q(M)\) rotates
all electrons simultaneously.

Numerically, Eq.~\eqref{eq:L0_projection} is approximated by a weighted
\(SO(3)\) quadrature summation,
\begin{equation}
\psi_{L=0}^{\rm net}
\simeq
\sum_{g=1}^{N_{\rm rot}}
\omega_g\,\hat U_Q(M_g)\psi^{\rm net},
\qquad
\sum_g\omega_g=1.
\label{eq:discrete_L0_projection}
\end{equation}



\section{Integration grids}
\label{sec:integration_grid}

We adopt the Euler-angle convention
\[
M(\alpha,\beta,\gamma)
=
M_z(\alpha)M_y(\beta)M_z(\gamma),
\]
with \(\alpha,\gamma\in[0,2\pi)\), \(\beta\in[0,\pi]\), and the
normalized Haar measure
\begin{equation}
dM=d\mu(M)
=
\frac{\sin\beta\,d\alpha\,d\beta\,d\gamma}{8\pi^2},
\qquad
\int_{SO(3)}dM=1.
\end{equation}
A quadrature rule \(\{M_q,w_q\}\) with \(\sum_qw_q=1\) is exact
through degree \(d\) if it reproduces the Haar integrals of every
Wigner matrix element \(D^\ell_{mn}\) with \(\ell\le d\).
Equivalently, its discrete Wigner moments must satisfy
\begin{equation}
\mathcal M^\ell_{mn}
=\sum_q w_qD^\ell_{mn}(M_q)
=
\begin{cases}
1, & \ell=0,\\
0, & 1\le\ell\le d,
\end{cases}
\qquad -\ell\le m,n\le\ell.
\label{eq:wigner_moment}
\end{equation}
Here \(d\) denotes the quadrature degree. This condition preserves the singlet
component and eliminates all non-singlet components within the
specified degree range.

\subsection{Euler-angle Gauss--Legendre product grid}
\label{sec:euler}

The Euler product rule combines equispaced nodes in the periodic
angles \(\alpha\) and \(\gamma\) with Gauss--Legendre quadrature in
\(x=\cos\beta\)~\cite{Khalid2016}. The periodic nodes are
\begin{equation}
\alpha_i=\frac{2\pi i}{N_\alpha},
\qquad
\gamma_j=\frac{2\pi j}{N_\gamma},
\end{equation}
where \(i=0,\ldots,N_\alpha-1\) and
\(j=0,\ldots,N_\gamma-1\).
The polar nodes \(x_k\), \(k=1,\ldots,N_\beta\), are the roots of
the Legendre polynomial \(P_{N_\beta}\), with
\(\beta_k=\arccos x_k\).

The associated Gauss--Legendre weights are
\begin{equation}
\lambda_k
=
\frac{2}
{(1-x_k^2)\left[P'_{N_\beta}(x_k)\right]^2},
\qquad
\sum_{k=1}^{N_\beta}\lambda_k=2,
\label{eq:gl_weights}
\end{equation}
where the prime denotes differentiation with respect to \(x\).
These weights approximate the integral over \([-1,1]\),
\[
\int_{-1}^{1}f(x)\,dx
\approx
\sum_{k=1}^{N_\beta}\lambda_k f(x_k),
\]
and the equality is exact for polynomials of degree at most
\(2N_\beta-1\). Combining the three one-dimensional rules gives
\begin{equation}
R_{ikj}
=
R_z(\alpha_i)R_y(\beta_k)R_z(\gamma_j),
\qquad
W_{ikj}
=
\frac{\lambda_k}{2N_\alpha N_\gamma}.
\end{equation}

Using
\[
D^\ell_{mn}(\alpha,\beta,\gamma)
=
e^{-\mathrm{i}m\alpha}
d^\ell_{mn}(\beta)
e^{-\mathrm{i}n\gamma},
\]
the discrete Fourier sums eliminate all nonzero \(m\) and \(n\)
within the range
\(\ell\le\min(N_\alpha-1,N_\gamma-1)\).
The surviving term is
\(d^\ell_{00}(\beta)=P_\ell(\cos\beta)\), which is integrated by
the Gauss--Legendre rule. The product grid is therefore guaranteed
to be exact through
\begin{equation}
d_{\mathrm{avg}}
=
\min(N_\alpha-1,N_\gamma-1,2N_\beta-1).
\end{equation}
For a requested degree \(d\), a sufficient choice is
\begin{equation}
N_\alpha=N_\gamma=d+1,
\qquad
N_\beta=\left\lceil\frac{d+1}{2}\right\rceil,
\end{equation}
giving the node count
\begin{equation}
N_{\mathrm{Euler}}^{SO(3)}(d)
=
(d+1)^2
\left\lceil\frac{d+1}{2}\right\rceil.
\end{equation}

\subsection{Lebedev quadrature lifted to \(SO(3)\)}
\label{sec:lebedev_gamma}

A Lebedev rule of degree \(p\) on \(S^2\) consists of nodes
\(\Omega_i=(\theta_i,\phi_i)\) and weights \(w_i\), normalized as
\(\sum_iw_i=4\pi\), and integrates spherical harmonics through
degree \(p\)~\cite{Lebedev1999}. To obtain an \(SO(3)\) rule, we
attach an equispaced \(\gamma\) grid to each spherical node:
\begin{equation}
R_{ij}
=
R_z(\phi_i)R_y(\theta_i)R_z(\gamma_j),
\qquad
\gamma_j=\frac{2\pi j}{N_\gamma},
\qquad
W_{ij}=\frac{w_i}{4\pi N_\gamma}.
\end{equation}

For \(N_\gamma=d+1\), the \(\gamma\) sum removes every nonzero
\(n\) mode with \(|n|\le d\). The remaining functions
\(D^\ell_{m0}(\phi_i,\theta_i,0)\) are proportional to spherical
harmonics and are integrated exactly when \(p\ge d\).

Let \(p(d)\) be the smallest available Lebedev degree not smaller
than \(d\), and let \(N_{\mathrm{Leb}}(p)\) denote its number of
spherical nodes. The lifted rule then contains
\begin{equation}
N_{\mathrm{Leb}+\gamma}(d)
=
N_{\mathrm{Leb}}\bigl(p(d)\bigr)(d+1)
\end{equation}
rotations. Specifying \(p(d)\) is necessary because standard Lebedev
rules are available only at selected degrees.

\subsection{Gauss-type \(SO(3)\) quadrature}
\label{sec:gauss_so3}

Rather than combining separate angular grids, Gauss-type rules
optimize the nodes and weights directly on \(SO(3)\), possibly
exploiting finite-group symmetries~\cite{GrafPotts}. We use selected
positive-weight rules from the published Chemnitz
collection.
The source reports numerically evaluated worst-case quadrature
errors below \(10^{-12}\) on the prescribed finite-degree space.

Supplementary Table~\ref{tab:grid_compare} compares the node counts of the three
constructions. If a Gauss-type rule is not tabulated at the requested
degree, a rule of higher degree can be used.

\begin{table}[htb]
\caption{\textbf{Node counts for \(SO(3)\) quadrature exact through
degree \(d\).}
The Euler and Lebedev\,+\,\(\gamma\) counts follow the constructions
in Supplementary Notes ~\ref{sec:euler} and \ref{sec:lebedev_gamma}.}
\centering
\begin{tabular}{cccc}
\toprule
\(d\) & Euler \(SO(3)\) & Lebedev\,+\,\(\gamma\)
      & Gauss-type \(SO(3)\) \\
\midrule
3  & 32   & 24   & 23 \\
4  & 75   & 70   & 43 \\
5  & 108  & 84   & 60 \\
6  & 196  & 182  & 116 \\
7  & 256  & 208  & 168 \\
8  & 405  & 342  & 240 \\
9  & 500  & 380  & 300 \\
10 & 726  & 550  & 504 \\
11 & 864  & 600  & 504 \\
12 & 1183 & 962  & 960 \\
14 & 1800 & 1290 & 960 \\
\bottomrule
\end{tabular}
\label{tab:grid_compare}
\end{table}

\section{Cusp}
\label{sec:si_cusp}
We simplify the derivation by taking $\ell_B=1$. The electron--electron cusp condition follows by canceling the
$1/r_{12}$ singularity of the Coulomb local energy at coalescence~\cite{Kato1957}.
For the two-electron Hamiltonian, we have
\[
H=-\tfrac12\nabla_{1}^{2}-\tfrac12\nabla_{2}^{2}
  +\frac{\kappa}{r_{12}}.
\]
Introducing the relative coordinate $\mathbf r=\mathbf r_1-\mathbf r_2$
gives
\[
H_{\mathrm{rel}}=-\nabla_\mathbf{r}^{2}+\frac{\kappa}{r}.
\]
In $D$ spatial dimensions, expand a relative wavefunction of angular
momentum $l$ as
\[
\Psi_{\mathrm{rel}}(\mathbf r)
=r^{l}(1+b_l r)Y_l(\hat{\mathbf r}),\qquad
\Delta_{S^{D-1}}Y_l=-l(l+D-2)Y_l.
\]
The leading terms of the Laplacian and local energy are
\[
\nabla_{\mathbf{r}^2}\Psi_{\mathrm{rel}}
=b_l(2l+D-1)r^{l-1}Y_l+O(r^l),
\]
\[
E_{\mathrm{loc}}
=\frac{\kappa-b_l(2l+D-1)}{r}+O(1),
\]
so the Coulomb singularity cancels when
\[
b_l=\frac{\kappa}{2l+D-1}.
\]
For fully spin-polarized electrons, the lowest allowed relative angular
momentum is $l=1$, and the local dimension is $D=2$, hence
$b_1=\kappa/3$. Since the Jastrow uses the unit-sphere chord
distance $\rho_{ij}=r_{ij}/R$, the corresponding dimensionless
coefficient is $c=b_1R=\kappa R/3$,
consistent with Eq.~\eqref{eq:pade_jastrow}.

%
\section{Kinetic estimator}
On the Haldane sphere of radius $R$ with monopole strength $Q$, the dimensionless kinetic operator is given by
\begin{equation}
\hat\Lambda_i^{2}
=-\frac{1}{\sin\theta_i}\partial_{\theta_i}
 (\sin\theta_i\,\partial_{\theta_i})
+\left(Q\cot\theta_i+\frac{\mathrm{i}}{\sin\theta_i}
\partial_{\varphi_i}\right)^2.
\label{eq:si_lambda_spherical}
\end{equation}
Direct differentiation of this expression is numerically unstable near
the north poles, because $1/\sin\theta_i$ and $\cot\theta_i$ become large.
We instead differentiate the wavefunction with respect to spinors.

Define the dimensionless single-particle angular momentum by
\begin{equation}
\hat{\boldsymbol L}_i=\hat{\boldsymbol\Lambda}_i+Q\mathbf n_i,
\qquad [\hat L_{ia},\hat L_{ib}]
=\mathrm{i}\epsilon_{abc}\hat L_{ic},
\qquad \hat{\boldsymbol L}_i^2=\hat{\boldsymbol\Lambda}_i^2+Q^2.
\label{eq:si_single_particle_L}
\end{equation}
The total kinetic-energy operator is therefore
\begin{equation}
\hat T=\frac{\hbar^2}{2mR^2}
\sum_{i=1}^N(\hat{\boldsymbol L}_i^2-Q^2).
\label{eq:casimir}
\end{equation}
For $a\in\{x,y,z\}$, we define
\begin{equation}
\begin{aligned}
G_a(\varepsilon)&=\exp\!\left(\frac{\mathrm{i}\varepsilon}{2}
\sigma_a^*\right),\qquad
\hat U_{i,a}(\varepsilon)=\exp(-\mathrm{i}\varepsilon\hat L_{ia}),\\
F_{ia}(\varepsilon;X)&=
\psi^{\rm net}(z_1,\ldots,G_a(\varepsilon)^{-1}z_i,
\ldots,z_N)
=\hat U_{i,a}(\varepsilon)\psi^{\rm net}(X).
\end{aligned}
\label{eq:si_partial_spinor_rotation}
\end{equation}
It follows that
\begin{equation}
\hat L_{ia}\psi^{\rm net}(X)
=\mathrm{i}\left.\frac{d F_{ia}}{d\varepsilon}
\right|_{\epsilon=0},\qquad
\hat L_{ia}^2\psi^{\rm net}(X)
=-\left.\frac{d^2F_{ia}}{d\varepsilon^2}
\right|_{\epsilon=0}.
\end{equation}

Since $(\mathbf n_i\cdot\boldsymbol\sigma^*)z_i=z_i$ and $\psi^{\rm net}$ is
homogeneous of degree $2Q$ in $z_i$, a rotation about the radial direction
$\mathbf n_i$ merely shifts the spinor phase and gives
$\hat L_{i,\mathbf n_i}^2\psi^{\rm net}=Q^2\psi^{\rm net}$, which cancels the
$-Q^2$ term in Eq.~\eqref{eq:casimir}. Thus only two rotations along an
orthonormal tangent pair $(\mathbf e_{i1},\mathbf e_{i2})\perp\mathbf n_i$ are
needed for each electron, $\hat{\boldsymbol\Lambda}_i^2=\hat L_{i,\mathbf e_{i1}}^2
+\hat L_{i,\mathbf e_{i2}}^2$.

\renewcommand{\refname}{Supplementary References}
\bibliography{supplement}